\documentclass[a4paper,11pt]{article}

\usepackage{jheppub} 

\usepackage[T1]{fontenc} 
\usepackage{CJK}
\usepackage{setspace}
\usepackage{float}
\def\be{\begin{equation}}
\def\ee{\end{equation}}
\def\bea{\begin{array}}
\def\eea{\end{array}}
\def\beqa{\begin{eqnarray}}
\def\eeqa{\end{eqnarray}}
\def\beqas{\begin{eqnarray*}}
\def\eeqas{\end{eqnarray*}}

\def\bp{\begin{picture}}
\def\ep{\end{picture}}
\def\bc{\begin{center}}
\def\ec{\end{center}}
\def\bfig{\begin{figure}}
\def\efig{\end{figure}}

\def\bit{\begin{itemize}}
\def\eit{\end{itemize}}
\def\nn{\nonumber}
\def\f{\frac}

\def\[{\left[}
\def\]{\right]}
\def\({\left(}
\def\){\right)}

\def\..{\left.}
\def\.{\right.}
\def\tl{\tilde}
\def\ra{\rightarrow}
\def\la{\leftarrow}

\def\tm{\times}

\def\da{\dagger}

\def\la{\lambda}

\def\al{\alpha}

\def\ka{\kappa}

\def\si{\sigma}
\def\ep{\epsilon}

\def\ga{\gamma}
\def\pa{\partial}

\title{ Solving The Muon g-2 Anomaly Within The NMSSM From Generalized Deflected AMSB}
\author[a]{Xuyang Ning,}
\author[a,b]{Fei Wang,}

\affiliation[a]{School of Physics, Zhengzhou University, Zhengzhou 450000, P. R. China}
\affiliation[b]{CAS Key Laboratory of Theoretical Physics, Institute of Theoretical Physics,
                Chinese Academy of Sciences, Beijing 100190, P. R. China}

\emailAdd{feiwang@zzu.edu.cn}
\emailAdd{ningxuyang@gs.zzu.edu.cn}

\abstract{ We propose to realize the (natural) NMSSM spectrum from deflected AMSB with new messenger-matter interactions. With additional messenger-matter interactions involving ${\bf 10}\oplus{\bf \overline{10}}$ representation messengers, the muon g-2 anomaly can be solved at $2\sigma$ (or $3\sigma$) level with the corresponding gluino mass range $2.8~{\rm TeV}<m_{\tl{g}}<5.4~ {\rm TeV}$ (or $2.6 ~{\rm TeV}<m_{\tl{g}}<7.3~ {\rm TeV}$). Besides, our scenario is fairly natural within which the involved fine tuning can be as low as 47. So, in the framework of AMSB-type scenarios, the NMSSM can be advantageous to explain the muon g-2 anomaly in compare with the MSSM.
}

\begin{document}
\maketitle \indent
\newpage
\section{Introduction}
Low energy supersymmetric (SUSY) theories, which was regarded for a long time as one of the most appealing extensions of the Standard Model(SM),  are well motivated to understand the large hierarchy between the electroweak(EW) scale and the Planck scale. The observed 125 GeV scalar discovered by both the ATLAS\cite{ATLAS:higgs} and CMS collaborations\cite{CMS:higgs} of LHC, although be compatible with a SM Higgs boson within experimental and theoretical uncertainties, can also be interpreted as the CP-even Higgs boson of the minimal supersymmetric standard model (MSSM). Besides, there are many other hints of low energy SUSY: the observed Higgs mass falls miraculously within the narrow $115-135$ GeV $'window'$ required by the MSSM; the mass of top quark mass also lies within the range required by SUSY to radiatively break the EW gauge symmetry; gauge coupling unification, which is not exact in SM, can be successfully realized in the framework of low energy SUSY; the lightest SUSY particle (LSP), which is stable under R-parity, can serve as a natural dark matter candidate with the right order of DM relic density.

It is well known that, to achieve the 125 GeV Higgs in the MSSM, we need either top squarks of order 10 TeV with small mixing or TeV-scale top squarks
with large mixing, reintroducing some amount of fine-tuning in the MSSM. This difficulty can be ameliorated in the framework of the Next-to-Minimal Supersymmetric Standard Model(NMSSM). The NMSSM, which is the simplest gauge singlet extension of the MSSM\cite{nmssm}, has various preferable features and had drawn a lot of attention in recent years. It can elegantly solve the $\mu$-problem in the MSSM by generating an effective $\mu$-term once the singlet scalar acquires a vacuum expectation value (VEV). Furthermore, due to possible new tree level contributions to the Higgs mass, the NMSSM can easily accommodate the 125 GeV Higgs boson, ameliorating the fine-tuning required.

 However, the lack of significant hints in EW precision tests and the null observation of conclusive signals for superparticles at the LHC
seem to make the possible discovery of weak scale SUSY doubtable. Recent analysis by LHC have set rather strong constraints on the sparticle  masses within the context of various simplified SUSY models: the gluino  $m_{\tl{g}} >1.5\sim 1.9$ TeV\cite{atlas:gluino} and the lighter top squark  $m_{\tl{t}_1} > 0.85$ TeV\cite{atlas:stop} with even stronger limits on the first generation squarks. Given the rapid accumulation of luminosity by LHC, a 2 TeV gluino is expected to be probed in some SUSY model. Besides, in order to account for the anomaly of muon anomalous magnetic moment in Brookheaven experiments, the weak scale SUSY spectrum should display an intricate structure. More than 3$\sigma$ discrepancy between the SM prediction\cite{sm:g-2} and the E821 measurement on\cite{e821} $g_\mu-2$ requires the existence of relatively light sleptons and EW gaugino\cite{mssm:g-2}. So supersymmetry breaking mechanism, which determines the whole low energy SUSY spectrum, is thus very crucial to understand such non-generic spectrum.

There are many interesting ways to mediate the SUSY breaking effects in the hidden sector to the visible MSSM sector. The anomaly mediation\cite{AMSB} SUSY breaking mechanisms (AMSB), which is determined by one parameter $F_\phi\simeq m_{3/2}$, is insensitive to the UV theory\cite{AMSB:RGE} and predicts a flavor conservation soft SUSY breaking spectrum.
 Unfortunately, the minimal AMSB predicts negative slepton masses and must be extended. The most elegant solution to tachyonic slepton is the deflected  AMSB\cite{deflect} scenario in which additional messenger sectors are introduced to deflect the AMSB trajectory and lead to positive slepton mass with additional gauge mediation contribution. On the other hand, $N\geq 4$ species are needed to give positive slepton masses and 125 GeV Higgs with naturally small negative deflection parameters, possibly leading to strong gauge couplings below GUT scale (or Landau pole below Planck scale). So less messenger species are preferable in deflect AMSB. We propose a way to reduce the number of messenger species by introducing general messenger-matter interactions in deflected
AMSB\cite{Fei:1508.01299} with small deflection parameter. The low energy NMSSM from typical SUSY breaking mechanism, such as GMSB, is always bothered by the requirement to achieve successful EWSB with suppressed trilinear couplings $A_\ka,A_\la$ and $m_S^2$, rendering the model building non-trivial\cite{NMSSM:GMSB}. Such difficulty can be ameliorated in AMSB-type scenarios with enhanced trilinear couplings. We find that the phenomenological interesting NMSSM spectrum can be successfully generated with generalized deflected AMSB mechanism involving messenger-matter interactions.

This paper is organized as follows. In Sec 2, we propose our model and discuss the general expressions for soft SUSY parameters from deflected AMSB with general messenger-matter interactions.
 Relevant numerical results are studied in Sec 3. Sec 4 contains our conclusions.

\section{The NMSSM From Deflected AMSB With Messenger-Matter Interactions}
It is known that standard GMSB or AMSB can not give a viable NMSSM spectrum. Lacking gauge interactions for $S$, the $A_\la, A_\ka$-terms are typically very suppressed in GMSB. In AMSB, large $A_\la,A_\ka$ needs large $\la$ and $\ka$ so as to induce large positive $m^2_S$, suppressing the singlet VEV\cite{NMSSM:AMSB}. By introducing typical form of the messenger sector with messenger-matter interactions, negative $m^2_S$ and large $A_\la,A_\ka$ can be generated in our scenario.

The framework of the NMSSM includes an extra gauge singlet superfield $S$ in addition to the MSSM sector. In the $Z_3$-invariant NMSSM, we have a scale invariant superpotential without the linear, bilinear terms for $S$  and the explicit $\mu$-term. Superpotential of the $Z_3$-invariant NMSSM is given by
\beqa
W_{NMSSM}&=&W_{MSSM}|_{\mu=0}+\la S H_u H_d+\f{\ka}{3} S^3~,~
\eeqa
with
\beqa
W_{MSSM}|_{\mu=0}=y^u_{ij} Q_{L,i} H_u U_{L,j}^c- y^d_{ij} Q_{L,i} H_d D_{L,j}^c-y^e_{ij} L_{L,i} H_d E_{L,j}^c~.
\eeqa
The soft SUSY breaking parameters are given as
\beqa
{\cal L}_{\tiny Z_3NMSSM}^{soft}={\cal L}_{MSSM}^{soft}|_{B=0}-\(A_\la \la S H_u H_d+A_\ka \f{\ka}{3} S^3\)-m_S^2 |S|^2~.
\eeqa

 We propose to generate the soft SUSY breaking parameters of the NMSSM in the framework of deflected AMSB mechanism.
The form of the superpotential at the GUT scale (upon the messenger threshold) includes the messenger sectors and superpotential $W(X)$ for pseudo-moduli $'X'$ with
\beqa
W=W_{\tiny Z_3NMSSM}+\sum\limits_{i}\[\la_X X\bar{Q}_i Q_i + \sum\limits_{a}\la_S S {\bf 10}_a \bar{Q}_i\]+ W(X)~.
\eeqa
Here we introduce $'N'$ family of vector-like messengers in the ${\bf 10}\oplus {\bf \overline{10}}$  representations of $SU(5)$ GUT group
\beqa
&&Q_i({\bf 10})=TQ_i(3,2)_{1/3}\oplus TU_i(\bar{3},1)_{-4/3} \oplus TE_i(1,1)_{2}~,\nn\\
&&\bar{Q}_i({\bf \overline{10}})=\overline{TQ}_i(\bar{3},2)_{-1/3}\oplus\overline{TU}_i(3,1)_{4/3}\oplus \overline{TE}_i(1,1)_{-2}~.
\eeqa
 Vector-like ${\bf 5}\oplus {\bf \bar{5}}$ messengers can also be possibly introduced. We fit the standard model matter contents in the $\bar{\bf 5}_a$ and ${\bf 10}_a$ representation of SU(5). In the superpotential, the messenger-matter interactions for $\tl{E}_{L,a}^c$ sleptons are introduced to give new contributions to the soft masses for such sleptons. Note that it is also possible to introduce the interactions for $\tl{L}_{L;a}$ (with interactions for ${\bf \bar{5}}$ messengers) or both. Possible massive messenger fields can be included in the superpotential without affecting the AMSB trajectory upon the threshold determined by the pseudo-moduli. It is well known that possible domain wall problems will arise in the $Z_3$-symmetric NMSSM. We assume that the $Z_3$ symmetry is broken by some higher dimensional operators.

 Deflection from AMSB trajectory can be used to solve the tachyonic slepton problem. The deflection parameter, which characterize the deviation from the ordinary AMSB trajectory, will depend on the form of the pseudo-moduli superpotential $W(X)$. Such superpotential can be fairly generic and lead to a deflection parameter $'d'$ of either sign given by
\beqa
d F_\phi\equiv \f{F_X}{X}-F_\phi~.
\eeqa
  In many circumstances, positive deflection parameter\cite{okada} could be welcome since less messenger species are needed. Positive deflection parameter in AMSB can be realized by a carefully chosen superpotential\cite{okada} or from strong dynamics\cite{Fei:1505.02785}.

  After Renormalization Group Equation(RGE) running to the messenger scale set by the pseudo-moduli VEV, the messenger sector reduces
\beqa
W_{M}&=&\sum\limits_{i}\[\la_X^Q X\overline{TQ}_i {TQ}_i+\la_X^U X\overline{TU}_i {TU}_i+\la_X^E X\overline{TE}_i {TE}_i +
 \la_{Q,a}^i S Q_{L,a} \overline{TQ}_i~\right.\nn\\&&~~~~+\left. \la_{U,a}^i S U_{L,a}^c \overline{TU}_i
+\la_{E,a}^i S (E_{L,a}^c) \overline{TE}_i \]+ W(X).
\eeqa
For simply, we adopt the simplest case with one family of $\overline{\bf 10}\oplus{\bf 10}$ messengers in our scenario. It is known that one additional vector-like {\bf 10} representation messengers will not spoil perturbative gauge coupling unification. We also adopt the case with $'a=1,2,3'$ which means that the messenger-matter interactions exist for all the three generations. The messenger-matter interactions involving the gauge singlet $S$ could change its AMSB trajectory and at the same time avoid possible mixing between $S$ and $X$.

The soft SUSY breaking parameters in deflected AMSB can be given by a typical mixed anomaly and gauge mediation contribution.
After RGE running down below the messenger threshold, the anomaly part will receive threshold corrections after integrating
out the messenger fields. We use the wavefunction renormalization approach\cite{wavefunction:hep-ph/9706540,chacko} and replace the messenger threshold $M_{mess}$ and the RGE scale $\mu$ by spurious chiral fields $X$ with
\beqa
 M_{mess}\ra \sqrt{\f{X^\da X}{\phi^\da\phi}}~,~~~~~\mu\ra \f{\mu}{\sqrt{\phi^\da\phi}} ~,
\eeqa
in the wavefunction superfield and gauge kinetic superfield to obtain the soft SUSY parameters. We reformulate some of the previous results below\cite{Fei:1508.01299, Fei:1602.01699,Fei:1703.10894}.
 The soft gaugino mass is given at the messenger scale by
\beqa
M_{i}(M_{mess})&=& g_i^2\(\f{F_\phi}{2}\f{\pa}{\pa \ln\mu}-\f{d F_\phi}{2}\f{\pa}{\pa \ln |X|}\)\f{1}{g_i^2}(\mu,|X|,T)~,
\eeqa
with
\beqa
\f{\pa}{\pa \ln |X|} g_i(\al; |X|)=\f{\Delta b_i}{16\pi^2} g_i^3~,
\eeqa

 The trilinear soft terms will be determined by the superpotential after replacing canonical normalized superfields and given by
\beqa
A_0^{ijk}\equiv \f{A_{ijk}}{y_{ijk}}&=&\sum\limits_{i}\(-\f{F_\phi}{2}\f{\pa}{\pa\ln\mu}+{d F_\phi}\f{\pa}{\pa\ln X}\) \ln \[Z_i(\mu,X,T)\]~,\nn\\
&=&\sum\limits_{i} \(-\f{F_\phi}{2} G_i^- +d F_\phi\f{\Delta G_i}{2}\)~.
\eeqa
In our convention, the anomalous dimension are expressed in the holomorphic basis\cite{shih}
\beqa
G^i\equiv \f{d Z_{ij}}{d\ln\mu}\equiv-\f{1}{8\pi^2}\(\f{1}{2}d_{kl}^i\la^*_{ikl}\la_{jmn}Z_{km}^{-1*}Z_{ln}^{-1*}-2c_r^iZ_{ij}g_r^2\).
\eeqa
with $\Delta G\equiv G^+-G^-$ the discontinuity across the messenger threshold. Here $G^+(G^-)$ denote respectively the value upon (below) the messenger threshold.

The soft scalar masses are given by
\beqa
m^2_{soft}&=&-\left|-\f{F_\phi}{2}\f{\pa}{\pa\ln\mu}+d F_\phi\f{\pa}{\pa\ln X}\right|^2 \ln \[Z_i(\mu,X,T)\]~,\\
&=&-\(\f{F_\phi^2}{4}\f{\pa^2}{\pa (\ln\mu)^2}+\f{d^2F^2_\phi}{4}\f{\pa}{\pa(\ln |X|)^2}
-\f{d F^2_\phi}{2}\f{\pa^2}{\pa\ln|X|\pa\ln\mu}\) \ln \[Z_i(\mu,X,T)\],\nn
\eeqa
 Details of the expression involving the derivative of $\ln X$ can be found in our previous works\cite{Fei:1508.01299, Fei:1602.01699,Fei:1703.10894}.

  We can see that there are several origins of the contributions in addition to the AMSB contributions:
\bit
\item  The gauge-anomaly interference contributions:
\beqa
\delta^I&=&\f{\pa^2}{\pa \ln X \pa\ln\mu} Z_i^-=\f{\pa}{\pa \ln X} G_i^-(Z_a,g_m)~,\nn\\
&=&\[\f{\pa Z_a}{\pa \ln X}\f{\pa}{\pa Z_a}+\f{\pa g_m}{\pa \ln X}\f{\pa}{\pa g_m}\]G_i^-(Z_a,g_m)~,
\eeqa
with
\beqa
\f{\pa }{\pa \ln X} Z_a= \f{\Delta G_a}{2}~,~~~~~~\f{\pa}{\pa \ln X} g_i=\f{1}{2}\f{\Delta b_i}{16\pi^2}g_i^3.
\eeqa
\item  Pure gauge mediation contributions
\beqa
\delta^G=\f{\pa^2}{\pa \ln X^\da \pa\ln X} Z_i^-.
\eeqa
\eit

With additional interactions for the gauge singlet superfield $S$,
the soft SUSY breaking scalar masses $m_S^2$ and trilinear coupling $A_\ka,A_\la$ will receive new contributions
so as to ameliorate the apparent conflict between the requirements of negative $m_S^2$ and large $A_\ka,A_\la$ in ordinary AMSB.

From the previous analytic results for the general deflected AMSB scenario,
the soft SUSY breaking parameters in the NMSSM at the messenger scale after integrating out the messenger fields can be given explicitly.
The relevant tedious expressions can be found in the appendix.

\section{Numerical Results}

EWSB condition is one of the most important constraints for the top-down realization of the NMSSM spectrum. From the minimization conditions of the NMSSM Higgs potential, one of the EWSB condition in the NMSSM can be written as
  \beqa
 \f{M_Z^2}{2}&=&\f{m^2_{H_d}-m_{H_u}^2\tan\beta^2}{\tan^2\beta-1}-\mu_{eff}^2~.~~
\label{EWSB}
 \eeqa
   It is clear from such EWSB condition that $m_{H_u}^2,\mu_{eff}$ should be light and of order $M_Z^2/2$ to avoid large fine tuning.
On the other hand, the most important corrections $\Sigma_u^u$ to $m_{H_u}^2$ come from the one loop diagrams involving the top squarks and top quark.
 So, in order to guarantee small loop corrections to $m_{H_u}^2$, relatively light stops are needed to accommodate EW naturalness.
In the MSSM, light stops should excess 1 TeV (or even heavier without stop maximal mixing) to accommodate the observed 125 GeV Higgs.
In the NMSSM, large loop corrections involving stops are not necessary \cite{NMSSM:light-stop} for sizable $\la\gtrsim 0.5$ and low $\tan\beta$. The main constraints on light stops come from the LHC direct search constraints.

The messenger-matter interactions in our scenario are chosen to be universal for simply
\beqas
\la_{2,1}^Q=\la_{2,1}^U=\la_{2,1}^E=\la_0~,~~\la_{3}^Q=\la_{3}^U=\la_{3}^E=\la_1~,~~\la_X^Q=\la_X^U=\la_X^E=\la_2.
\eeqas
In principle, their precise values should be given by RGE running from the GUT scale to the messenger scale. However, the total messenger contents upon the messenger scale can be model dependent because of the $'decoupling~theorem'$\cite{AMSB:RGE} in  AMSB which state that simple messenger threshold (by pure mass term) will not deflect the AMSB trajectory at leading order. By assigning different origin for messenger thresholds (determined by moduli VEV or pure mass term), effects of certain messengers on low energy AMSB spectrum can decouple although they can still contribute to gauge coupling(Yukawa) unification. Because of the freedom to add messengers (with pure mass thresholds) upon messenger scale, we neglect the RGE effects for messenger-matter interactions in our numerical study for simply.

 The free parameters in our scenario are:
  \beqa
  d, M_{mess},F_\phi, \la,\ka,\la_0,\la_1,\la_2.
  \eeqa

  We need to check if successful EWSB condition is indeed fulfilled.
   In fact, the soft SUSY mass $m_{H_u}^2,m_{H_d}^2,m_S^2$ can be reformulated into $\mu,\tan\beta,M_Z^2$ by the minimum condition
   of the scalar potential. Usually, $M_A$ can be used to replace $A_\ka$ by
  \beqa
  M_A^2=\f{2\mu_{eff}}{\sin2\beta}B_{eff}~, ~~\mu_{eff}\equiv\la\langle s\rangle~,~~~B_{eff}=(A_\la+\ka \langle s\rangle).
  \eeqa
  We should note that $\kappa$ is a free parameter while $\tan\beta$ is not. This choice is different to ordinary numerical setting in the NMSSM in which $\tan\beta$ is free while $\kappa$ is a derived quantity\cite{ellwanger:0612134}. Our choice can be convenient for those predictable NMSSM models from top-down approach. A guess of $\tan\beta$ is made to obtain the relevant Yukawa $y_t,y_b$ couplings at the EW scale. After RGE evolving up to the messenger scale as the theory inputs, the whole soft SUSY breaking parameters at the messenger scale can be obtained. Low energy $\tan\beta$ can be obtained iteratively from such spectrum with minimization condition for the Higgs potential.

 The purpose of deflection in AMSB is to solve the notorious tachyonic slepton problem. So non-tachyonic slepton should be obtained at the SUSY scale.
 Positive slepton masses can be realized by introducing either sign of deflection parameter $d$ with proper matter-messenger interactions.

    We use the package NMSSMTools 5.01\cite{NMSSMTOOLS} to scan the whole parameter space. We will interest in relatively large values of $\la$ in order to increase the tree-level mass of the 125 GeV CP-even Higgs boson. The parameters are chosen to satisfy:
 \beqa
&&~~ 10^{6} GeV< M_{mess}< 10^{15} GeV~,~~ 0.5 {\rm TeV} <F_\phi< 100 {\rm TeV}~,~~-5<d<5,\nn\\
&&~~~~~~~~~~~0<\la_0,\la_1<\sqrt{4\pi}~,~~~~~~~~~~~~~~~~~0.1<\la,\ka <0.7~,
 \eeqa
Besides, we require that $\la^2+\ka^2\lesssim 0.7$ to satisfy the perturbative bounds.

 In our scan, we impose the following constraints:

\bit
\item  The lower bounds from current LHC constraints on SUSY particles\cite{atlas:gluino,atlas:stop}:
    \bit
    \item  Light stop mass: $m_{\tl{t}_1} \gtrsim 0.85$ TeV.
    \item  Gluino mass: $m_{\tl{g}} \gtrsim 1.5\sim 1.9 $ TeV.
    \item  Light sbottom mass $m_{\tl{b}_1} \gtrsim 0.84$ TeV.
    \item  Degenerated first two generation squarks $m_{\tl{q}} \gtrsim 1.0 \sim 1.4$ TeV.
    \eit
\item
     The lower bounds for neutralinos and charginos, including the invisible decay bounds for $Z$-boson.
     The most stringent constraints of LEP require $m_{\tl{\chi}^\pm}> 103.5 {\rm GeV}$ and
     the invisible decay width $\Gamma(Z\ra \tl{\chi}_0\tl{\chi}_0)<1.71~{\rm MeV}$,
     which is consistent with the $2\sigma$ precision EW measurement $\Gamma^{non-SM}_{inv}< 2.0~{\rm MeV}$.

\item  Flavor constraints from the rare decays of B mesons. We adopt the recent experimental results\cite{B-physics}:
  \beqa
  && 0.85\tm 10^{-4} < Br(B^+\ra \tau^+\nu) < 2.89\tm 10^{-4}~,\nn\\
  &&1.7\tm 10^{-9} < Br(B_s\ra \mu^+ \mu^-) < 4.5\tm 10^{-9}~,\nn\\
  && 2.99\tm 10^{-4} < Br(B_S\ra X_s \gamma) < 3.87\tm 10^{-4}~.
   \eeqa

\item  The CP-even component $S_2$ in the Goldstone-$'eaten'$ combination of $H_u$ and $H_d$ doublets corresponds to the SM Higgs.
 The $S_2$ dominated CP-even scalar should lie in the combined mass range for the Higgs boson: $122 {\rm GeV}<m_{h_0} <127 {\rm GeV}$ from ATLAS and CMS data. Note that the uncertainty is 3 GeV instead of default 2 GeV because large $\lambda$ may induce additional ${\cal O}(1)$ GeV correction to $m_{h_0}$ at two-loop level\cite{NMSSM:higgs2loop}, which is not included in the NMSSMTools.

\item The relic density of DM should satisfy the Planck data $\Omega_{DM} = 0.1199\pm 0.0027$ \cite{Planck}
in combination with the WMAP data \cite{WMAP}(with a $10\%$ theoretical uncertainty). In our scenario, only the upper bound for DM relic density is used.

\item The $g_\mu-2$ discrepancy should be solved in our scenario.
  The E821 experimental result on the muon anomalous magnetic moment at the Brookhaven AGS \cite{Mg-2:collaboration} is given as
 \beqa
a_\mu^{\rm expt} =116592089(63)\times 10^{-11}~,
 \eeqa
which is larger than the SM prediction
\beqa
a^{\rm SM}_\mu =116591834(49)\times 10^{-11}~.
\eeqa
   The deviation is about $3\sigma$
 \beqa
\Delta a_\mu({\rm expt - SM}) = (255\pm 80)\times 10^{-11}.
\eeqa
Recent discussions on $g_\mu-2$ anomaly can be seen in \cite{muon:g-2}.
\eit

\begin{figure}[htb]
\begin{center}
\includegraphics[width=5.0 in]{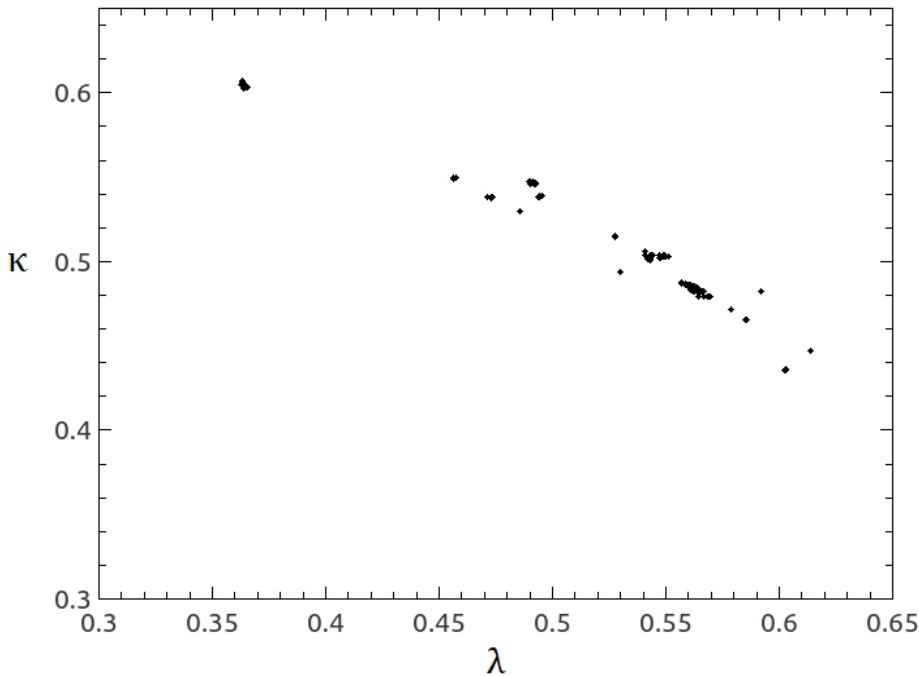}
\end{center}
\caption{The scattering plots of survived points for $\la$ versus $\ka$. All survived points satisfy the LHC and EW precision measurement constraints as well as the required value of $\Delta a_\mu$.}
\label{la-ka}
\end{figure}

We have the following discussions on our numerical results:
\bit
\item  Successful EWSB condition in the NMSSM always set stringent constraints on the constrained low energy inputs, especially when such low energy inputs are determined by certain UV-completed theory. In our scenario, the (deflected) AMSB-type mechanism with few free parameters determines the whole SUSY spectrum at the messenger scale. So successful EWSB condition will rule out much parameter space. Random scan in our scenario indicates that many points can still survive the EWSB conditions. The allowed range for the characteristic NMSSM parameters $\ka$ and $\la$ can be seen in fig.\ref{la-ka}. We can see that our numerical results give the lower bound for $\ka$ and $\la$ with $\ka_{\rm low}\approx 0.43$ and $\la_{\rm low}\approx 0.35$ in addition to the upper bounds $\ka,\la\sim 0.62$.

\begin{figure}[htb]
\begin{center}
\includegraphics[width=2.9in]{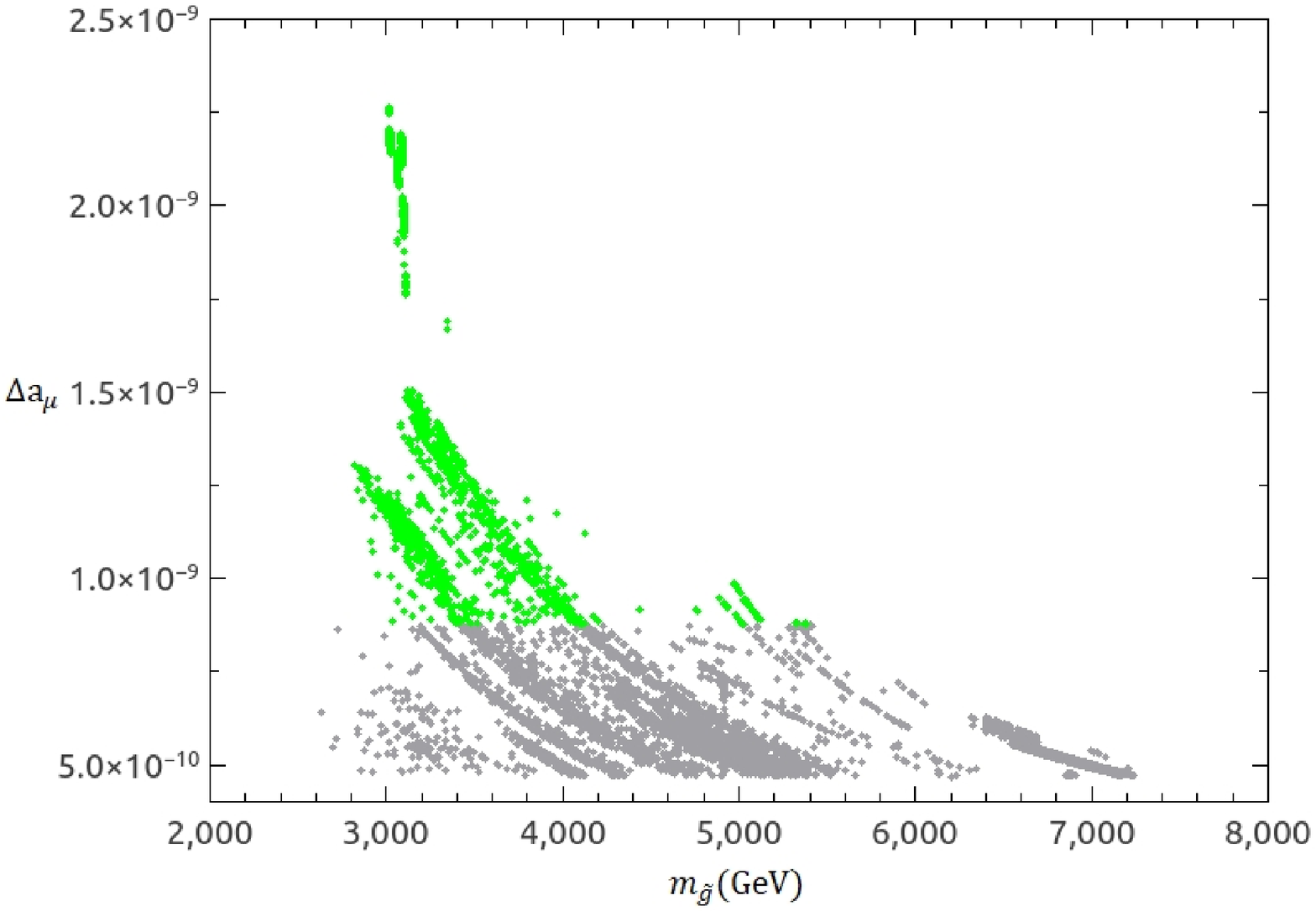}
\includegraphics[width=2.9in]{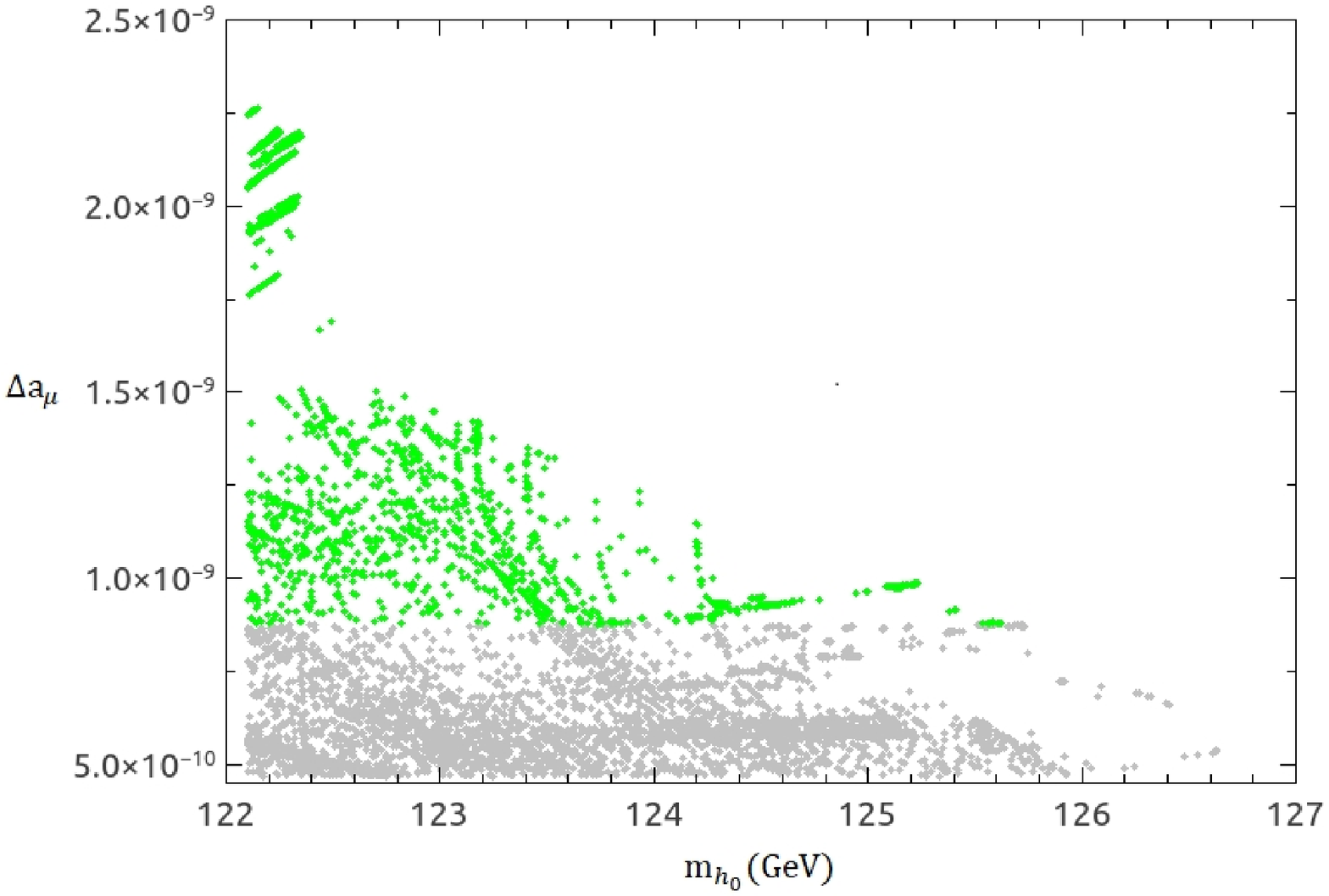}
\end{center}
\caption{The scattering plots of survived points for SUSY contributions to muon g-2 anomaly $\Delta a_\mu$ versus the gluino mass $m_{\tl{g}}$(left plot) and the SM-like Higgs mass $m_{h_0}$ (right plot). The green(gray) points satisfy the $2\sigma$($3\sigma$) range of the muon g-2 anomaly. All survived points satisfy the LHC and EW precision measurement constraints as well as the required value of $\Delta a_\mu$.}
\label{Deltag-G}
\end{figure}

\item  The left plot of fig.\ref{Deltag-G} shows the SUSY contributions to $g_\mu-2$ anomalous magnetic momentum $\Delta a_\mu$ versus the gluino mass $m_{\tl{g}}$.
It is obvious that the $g_\mu-2$ anomaly can be successfully solved at both the $2\sigma$(green points) and $3\sigma$ (grey points) level.
The MSSM is known to be able to yield sizable contributions to the $\Delta a_\mu$ which come dominantly from the chargino-sneutrino and the neutralino-smuon loops. The $g_\mu-2$ anomaly, of order $10^{-9}$, can be solved for $m_{\rm SUSY} = {\cal O}(100)$ GeV and $\tan\beta = {\cal O}(10)$ with positive $\mu$. Besides, such SUSY contributions increase with respect to $\tan\beta$. In our scenario, sleptons as well as $M_1,M_2$ can be relatively light while colored sparticles are heavy to evade possible constraints from LHC, SUSY flavor and CP problems.

\begin{figure}[htb]
\begin{center}
\includegraphics[width=2.9 in]{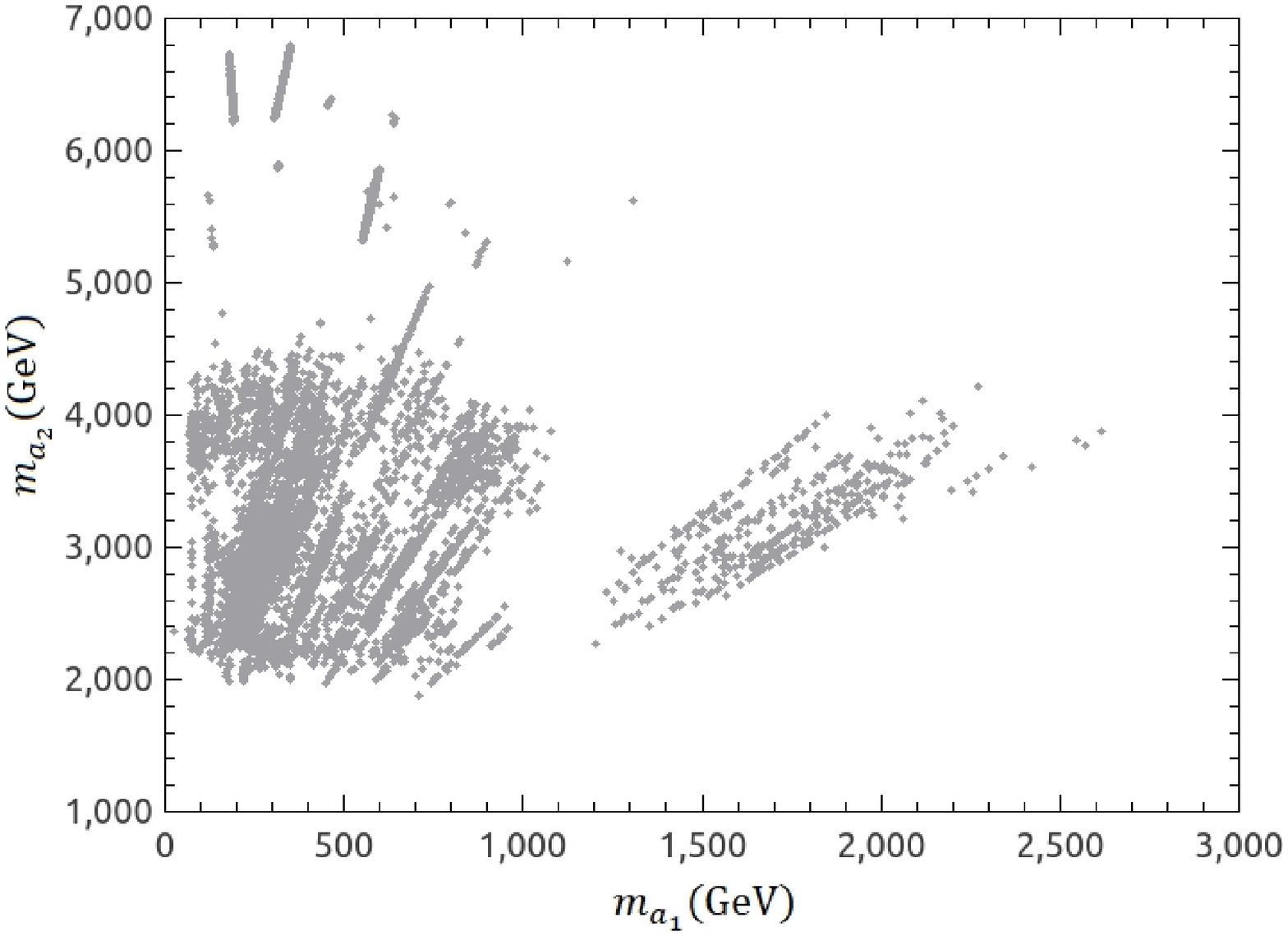}
\includegraphics[width=2.9 in]{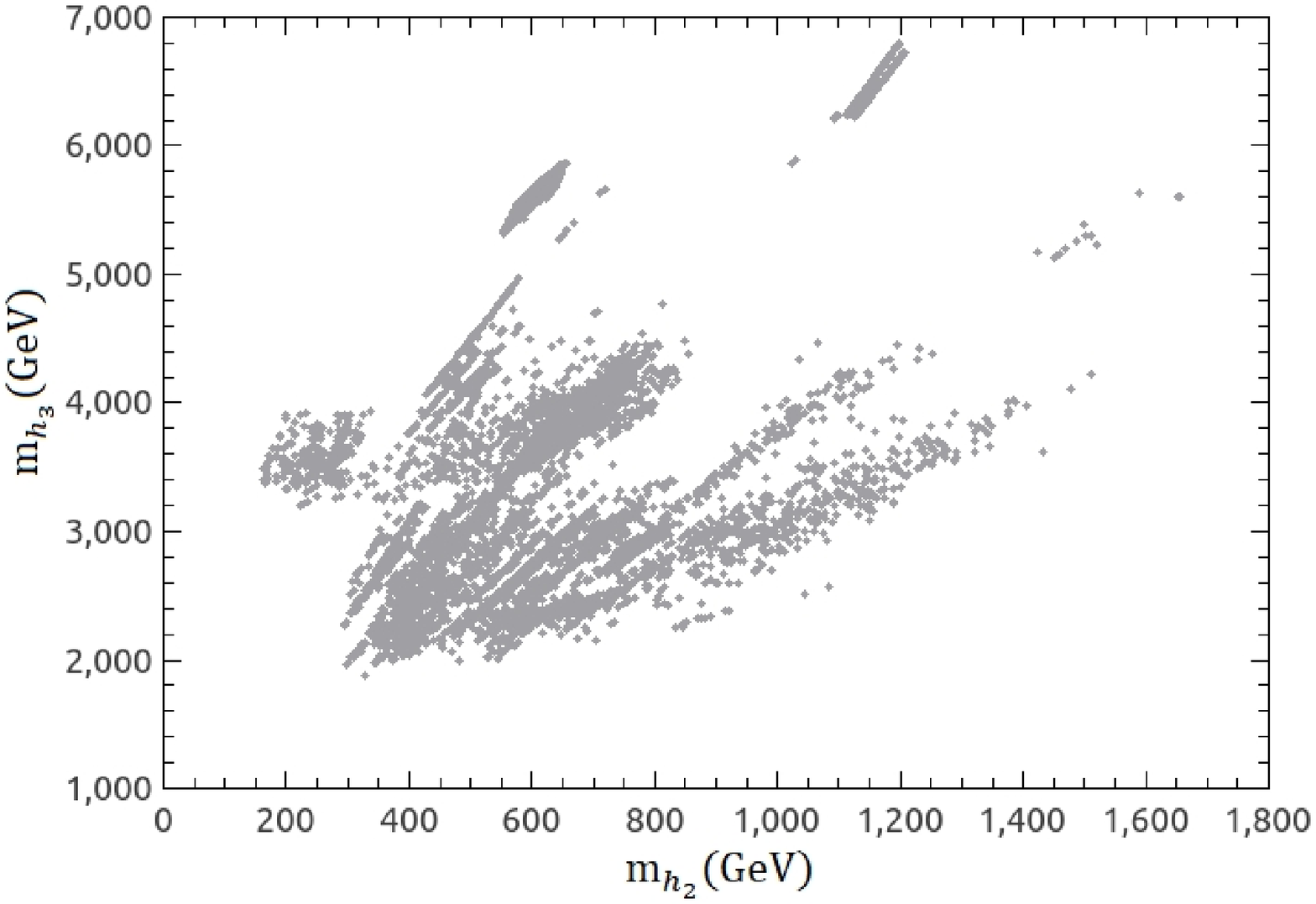}\\
\includegraphics[width=2.9in]{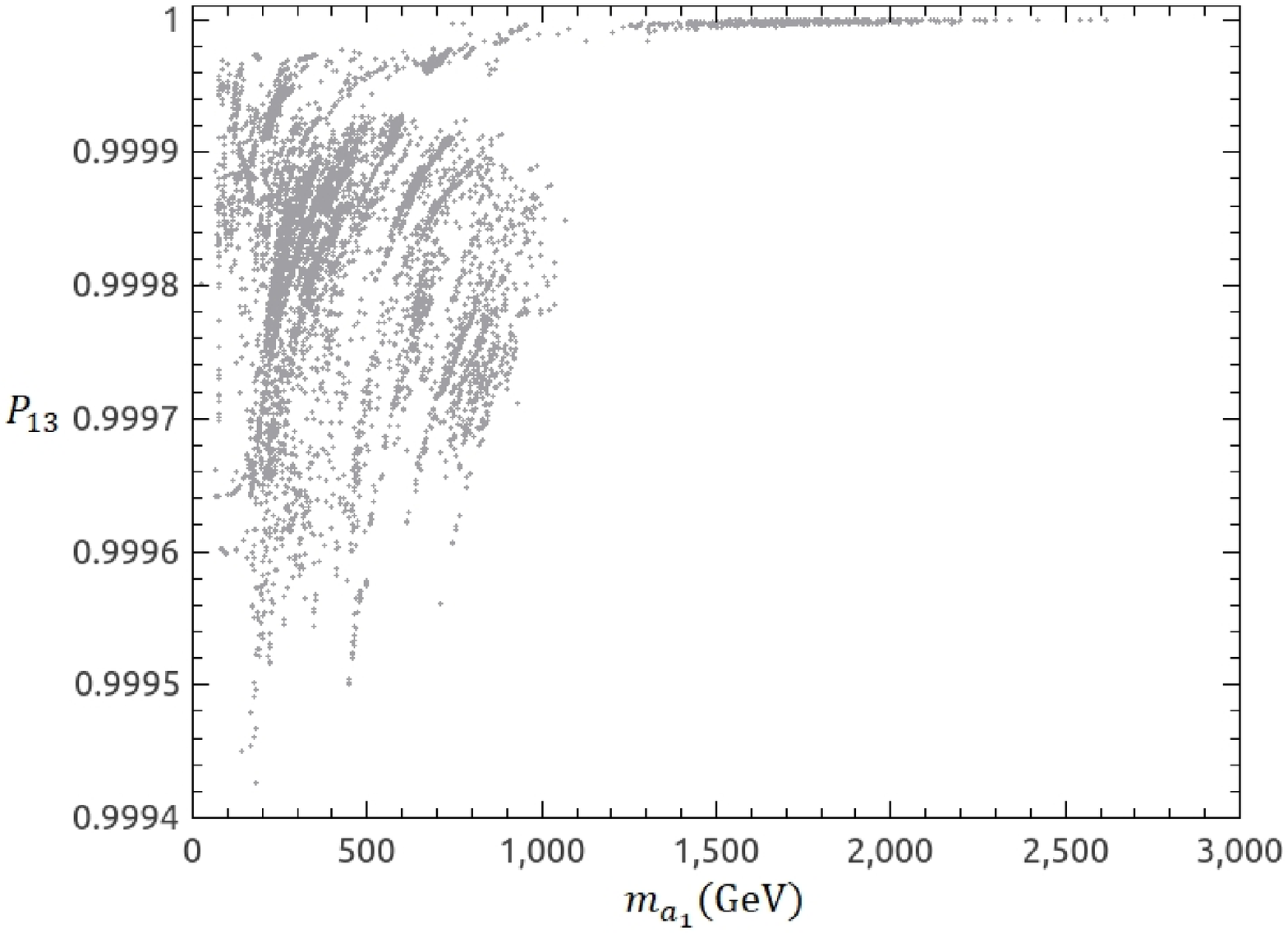}
\includegraphics[width=2.9in]{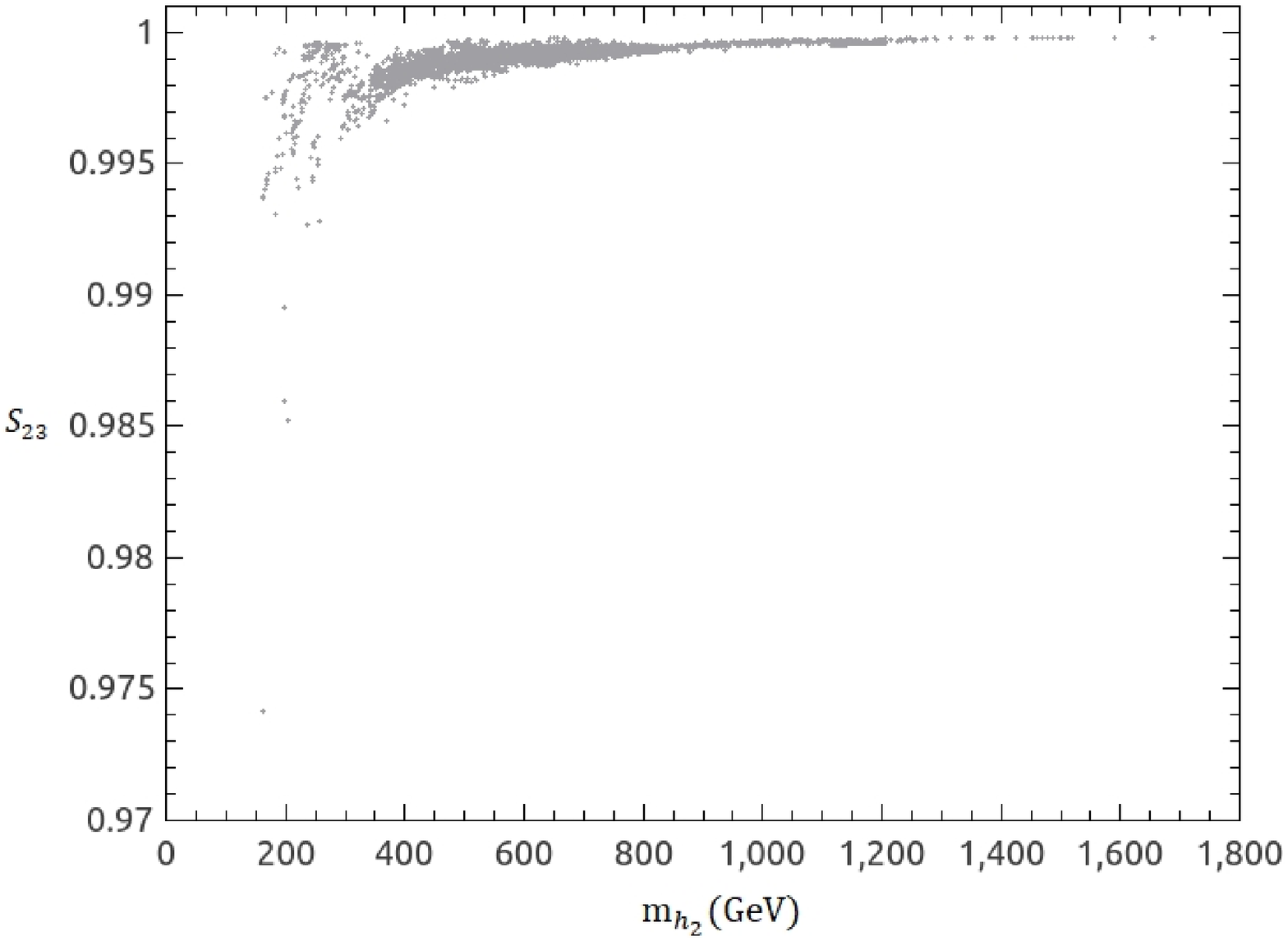}
\end{center}
\caption{ In the upper panels are the scattering plots of $m_{a_1}$ versus $m_{a_2}$(left) and $m_{h_2}$ versus $m_{h_3}$ (right). All survived points satisfy the LHC and EW precision measurement constraints as well as the required value of $\Delta a_\mu$ to solve the muon g-2 anomaly. In the lower panels are the plots of the singlet components for $a_1$ (left) and $H_2$ (right).}
\label{a1-a2}
\end{figure}

The inclusion of singlino in the NMSSM will not give sizable contributions to $\Delta a_\mu$ because of the suppressed coupling of singlino to the MSSM sector. Two loop contributions involving the Higgs is negligible in SM. In fact, the SM one-loop Higgs/muon diagram is, about four orders of magnitude below the sensitivity of the $g_\mu-2$ experiment. However, the new Higgs bosons could have an important impact on $a_\mu$ if the lightest neutral CP-odd Higgs scalar can be quite light\cite{NMSSM:g-2}. As noted in that paper, the positive two-loop contribution is numerically more important for a light CP-odd Higgs heavier than 3 GeV and the sum of both one loop and two loop contributions is maximal around $m_{a_1}\sim 6$ GeV. We show the CP-even and CP-odd Higgs masses in fig.\ref{a1-a2} for the allowed parameter space by $\Delta a_\mu$. From the left plot of fig.\ref{a1-a2}, we can see that the lightest CP-odd Higgs $a_1$, which always lies above 62 GeV, is not light enough to give sizable contributions to $\Delta a_\mu$. The main contributions are similar to that in the MSSM. We should also note that the 125 GeV SM-like Higgs always corresponds to the lightest CP-even Higgs boson for our survived points. The possibility that the 125 GeV Higgs being the second lightest CP-even scalar is not given in our scenario. The LHC discovery prospects for the scalar sector of the NMSSM can be seen in \cite{ellwanger:1512.04281,king:1408.1120,Bomark:2015fga,Djouadi:2008uw,Cao:2013gba}.

\item In most of the allowed parameter spaces, the lightest CP-odd scalar $a_1$ and the second lightest CP-even scalar $h_2$ are singlet dominated.
The corresponding singlet components of the $a_1$ and $h_2$ are displayed in the lower panels of fig.\ref{a1-a2}. If kinetically allowed, the $h_2$ scalar can decay primarily into $a_1$ scalar pairs and vector boson pairs.

 The cross section of $h_2$ production from dominant gluon fusion is small in compare with the SM-like Higgs because of its dominant singlet component.
In case the primary $h_{2}\ra a_1 a_1$ decay channel is open, the $W^+W^-,ZZ$ final states from gluon fusion produced $h_{2}$ scalar decay are typically suppressed by order $10^{-5}$ in compare with the SM-like Higgs of the same mass, making it difficult to give a signal of statistical significance at the LHC. On the other hand, in a small region of parameter space within which the $h_2\ra a_1 a_1$ channel is not open for light $h_2$ and heavy $a_1$, the $h_2$ can decay primarily into $W^+W^-$ pairs with
$ll\nu\nu(l=e,\mu),l\nu jj$ final states. Because of the reduced production rate and large background from $pp\ra W^+W^-$ and $t\bar{t}\ra b\bar{b}W^+W^-$, they could hardly be discovered by recent LHC experiments.
 The possibility to detect the singlet-dominated CP-even scalar in diphoton channel had been discussed in \cite{ellwanger:1512.04281}. However, in our scenario, the  branch ratio into diphoton is very small in compare with that for a SM Higgs boson of the same mass,
  making the diphoton final states not very promising to search for at $\sqrt{s}=13 {\rm TeV}$ LHC.

    From our numerical results, the CP-odd scalar $a_1$ should be heavier than 62 GeV (half of the SM-like Higgs mass). If this scalar is lighter than half of the SM-like Higgs boson, the SM-like Higgs boson could decay into $a_1$ pairs. Because the width of the SM-like Higgs boson is quite narrow, such an exotic decay may have a sizable branching ratio and in turn suppress greatly the visible signals of the SM-like Higgs boson at the LHC.
     So the LHC measurements, in particular those that are put on the Higgs exotic decay modes, will give stringent constraints on such light CP-odd scalar.

      The $a_1$ scalar can be produced from heavy scalar cascade decay or direct production in association with a bottom-quark pair\cite{moretti:1205.1683}. Each $a_1$ from $h_2\ra a_1a_1$ will decay primarily to $b \bar{b}$ and $\tau^+\tau^-$, yielding final states that will typically have large backgrounds at the LHC. Requiring one $a_1$ to decay to $\mu^+\mu^-$ pairs can allow one to use the dimuon invariant mass as a discrimination between the signal and the background. However, the $\sigma\times Br$ cross section is rather small. Possible $4\ga$ signal from both the diphoton decay modes of $a_1$ will not be significant enough due to the absence of a possible enhancement of the diphoton branching fraction. Two benchmark points are shown in Table \ref{AB}
      with the decay modes for light scalars and the $h_2$ production cross section from gluon fusion.






\begin{table}[h]
\caption{Two benchmark points with the corresponding branch ratio for $h_2,a_1$ and gluon fusion cross section for $h_2$ at LHC.}
\centering
\begin{tabular}{|c|c|c|}
\hline
Input & A & B  \\
\hline
$(\la, \ka,\tan\beta)$&(0.56, 0.48, 11.5)&(0.60, 0.43, 9.1)\\
\hline
$(m_{h_0},m_{h_2},m_{h_3})$&(123,~~674,~~3852)~{\rm GeV}&(123,~~162,~~3488)~{\rm GeV} \\
\hline
$(m_{a_1},m_{a_2},m_{\tl{\chi}_1^0})$& (63,~~3850,~~103)~{\rm GeV}&(929,~~3488,~~109)~{\rm GeV} \\
\hline
$Br(h_2\ra a_1 a_1; h_1h_1 ;\ga\ga)$& $(0.884 ;1.06\times10^{-2};1.23\times10^{-6})$&$(~\setminus ;~\setminus ;4.61\tm 10^{-4})$\\
\hline
$Br(h_2\ra W^+W^-;Z Z)$& $(3.64\tm 10^{-2};1.77\tm 10^{-2})$&$(0.966;2.65\tm 10^{-2})$ \\
\hline
$Br(a_1\ra \ga\ga)$& $(1.57\tm 10^{-4})$&$(1.36\tm 10^{-5},1.53\tm 10^{-5})$\\
\hline
$Br(a_1\ra b\bar{b}; \tau^+\tau^-)$& $(0.911;8.52\tm 10^{-2})$ &$(9.14\tm10^{-5};1.37\tm 10^{-5})$ \\
\hline
\hline
$\si(ggF\ra h_2)~{\rm @} \sqrt{s}=8;14 {\rm TeV}$ &(0.36; 1.44)~fb&(683; 1821)~fb\\
\hline
\end{tabular}
\label{AB}
\end{table}

\item
It is also clear from fig.\ref{Deltag-G} that the gluino should lie between 2.6 TeV to 7.3 TeV to solve the observed $g_\mu-2$ anomaly. The upper bound on gluino masses can be understood as follows: The scale of the low energy soft SUSY parameters from anomaly mediation is determined by the value of $F_\phi$. The lightness of $\tl{\mu},\tl{\nu}_\mu, \tl{B},\tl{W}$ to account for the $g_\mu-2$ anomaly sets an upper bound on $F_\phi$, which subsequently sets the mass scale of the gluino.

 The observed SM-like Higgs can also be explained in our scenario as anticipated (see the right plot of fig.\ref{Deltag-G}) because additional contributions from the NMSSM can increase the predicted SM-like Higgs mass. Besides, in order to solve the $g_\mu-2$ anomaly at $2\sigma(3\sigma)$ level, the Higgs mass can be as high as 125.6GeV (126.7 GeV). As a comparison, the Higgs mass is upper bounded by 118 GeV(120 GeV) if the $g_\mu-2$ anomaly is solved at $2\sigma(3\sigma)$ level in the CMSSM. So, our scenario is much better in solving the $g_\mu-2$ anomaly and satisfying the Higgs mass measurement.

  In the framework of (extended) anomaly mediation, the NMSSM seems to be advantageous in compare with the MSSM in which the gluino masses are bounded to below 2.0 TeV(2.5 TeV) if the $g_\mu-2$ anomaly is solved at $2\sigma$ ($3\sigma$) level with complete GUT multiplet messengers\cite{Fei:1703.10894}. In fact, the latest LHC search results had already set a lower bound $1.5\sim 1.9$ TeV on gluino mass. In most of the allowed parameter space, the MSSM from (deflected) AMSB can not solve the $g_\mu-2$ anomaly at $2\sigma$ level unless incomplete GUT multiplet messengers are introduced. Although the two scenarios in the MSSM and the NMSSM introduce different types of deflections in their messenger sectors, an important constraint on the parameters comes from the 125 GeV Higgs mass.

   In the MSSM, the observed 125GeV Higgs mass will set stringent constraints on the soft SUSY breaking parameters, especially on the trilinear coupling $A_t$ and stop masses $m_{\tl{t}_1},m_{\tl{t}_2}$. So the parameter $F_\phi\simeq m_{3/2}$, which determines the whole soft SUSY spectrum, is bounded to lie in a small region (in combing with other constraints). Thus, the gaugino masses, which are fully determined by $F_\phi$ and $d$, are constrained to lie below 2.5 TeV.
    On the other hand, the observed 125 GeV Higgs can easily be accommodated in the NMSSM and will not set very stringent constraints on the parameter $F_\phi$.
    So the gaugino masses, which are predicted by the value of $F_\phi$, can be relatively heavy in the NMSSM.

\begin{figure}[htb]
\begin{center}
\includegraphics[width=5.0 in]{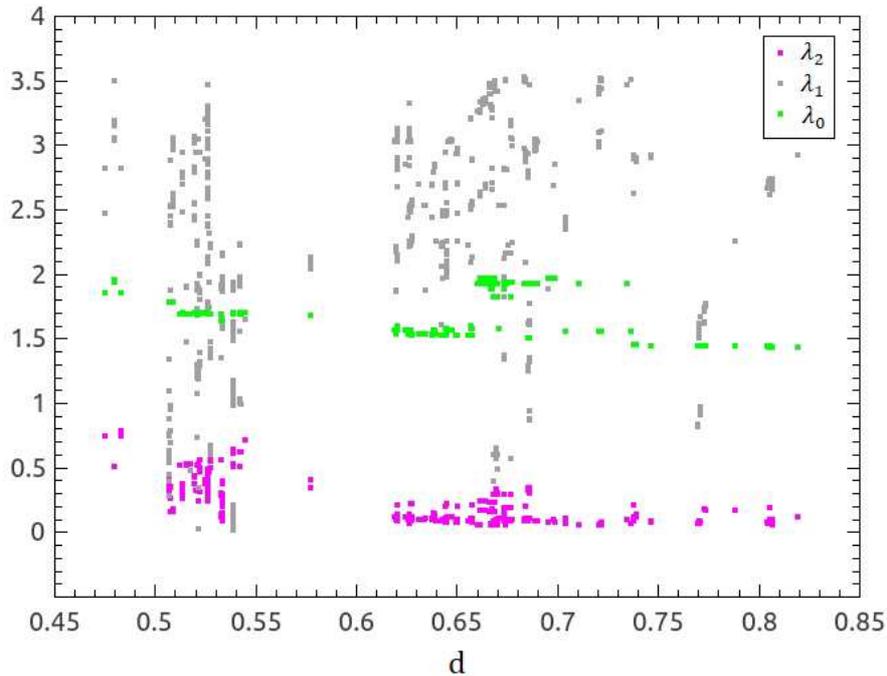}
\end{center}
\caption{The scattering plots of survived points for the deflection parameter $d$ versus the messenger-matter couplings $\la_{0,1,2}$. All survived points satisfy the LHC and EW precision measurement constraints as well as the required value of $\Delta a_\mu$.}
\label{d-la0}
\end{figure}

  The deflection with non-trivial messenger-matter interactions involving $\tl{\mu}_L^c$ is crucial in obtaining sizable contributions to $\Delta a_\mu$. We can see from fig.\ref{d-la0} that $1.4<\la_0<2$ with positive $'d'$ in the range $0.45<d<0.85$ is required for all allowed points in our scenario. Positive deflection in this range can cure the tachyonic slepton problem of ordinary AMSB and give small positive slepton masses to account for sizable $\Delta a_\mu$. Without such new messenger-matter interactions in AMSB type scenario, the two problems can not be solved simultaneously. More ${\bf 10\oplus \overline{10}}$ messenger species will change the allowed range of $d$. Similar conclusion holds for scenarios with messenger-matter interactions involving the doublet $(\tl{\nu}_{\mu;L},\tl{\mu}_L)$.

\item  As noted earlier, relatively light stops $m_{\tl{t}_1,\tl{t}_2}\lesssim 1.5$ TeV are needed by naturalness with EW fine tuning $\Delta_{EW}\gtrsim 10$ in the (radiative) natural MSSM\cite{rnaturalsusy}. Heavy stops are not necessarily required to interpret the 125 GeV Higgs in the NMSSM because of the additional tree-level contributions. So it seems natural to achieve (electroweak) naturalness in the NMSSM instead of the MSSM.
     Although new LHC search results had already imposed stringent constraints on the mass of top squarks to excess 1 TeV, EW naturalness with ${\cal O}$ (TeV) stops can still be possible with large $|A_t|$. Detailed studies on the radiative natural NMSSM will appear soon in our subsequent studies\cite{fei-cao}.

  The fine-tuning with respect to certain input parameter $a$ is defined as
          \beqa
          \Delta_a\equiv \left|\f{\pa \ln M_Z^2}{\pa \ln a}\right|~,
          \eeqa
 while the total fine-tuning is defined to be $ \Delta= \max\limits_{a}(\Delta_a)$ with $\{a\}$ the set of parameters defined at the input scale.

\begin{figure}[htb]
\begin{center}
\includegraphics[width=5.0 in]{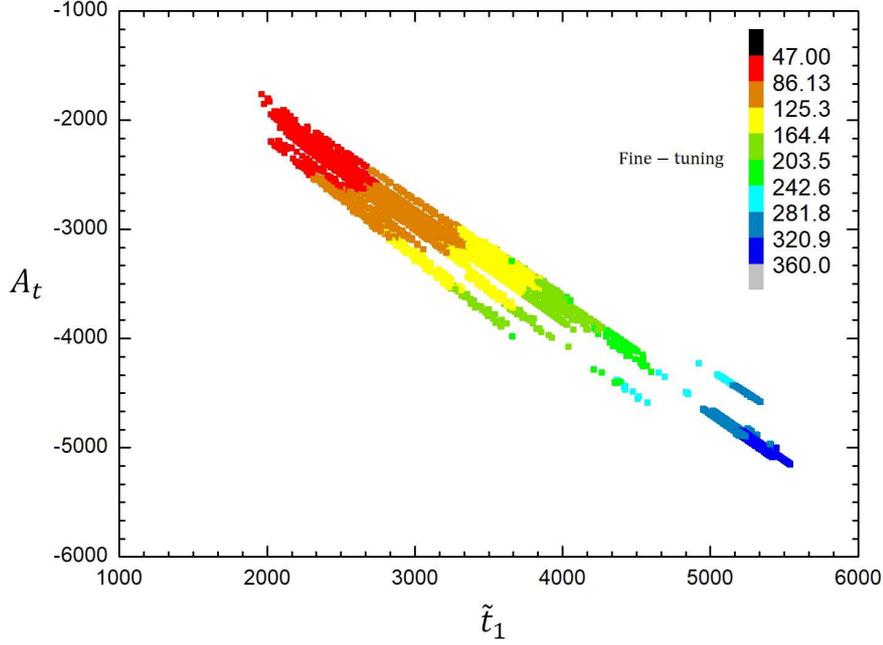}
\end{center}
\caption{Survived points with their lightest stop mass $m_{\tl{t}_1}$ versus the trilinear couplings for top Yukawa $A_t$ and also their corresponding fine tuning involved. Note that the trilinear coupling $A_t$ of all points are negative.}
\label{FT}
\end{figure}

 The fine-tuning of our scenario within the allowed parameter space by $\Delta a_\mu$, the LHC results and the EW precision measurement is shown in fig.\ref{FT}. In the allowed region, the fine tuning $\Delta\lesssim 360$. The fine tuning can be as low as 47 in certain regions which can be seen to generate a natural NMSSM spectrum. All the trilinear couplings $A_t$ are negative with $1.6 {\rm TeV}<m_{\tl{t}_1}<5.8 {\rm TeV}$.
As expected, the low fine-tuning region corresponds to light $\tl{t}_1$. Besides, for a given $A_t$, lighter ${\tl{t}_1}$ always corresponds to less fine tuning. Similar conclusion in the MSSM can be seen in \cite{rnaturalsusy}.

\begin{figure}[htb]
\begin{center}
\includegraphics[width=2.9 in]{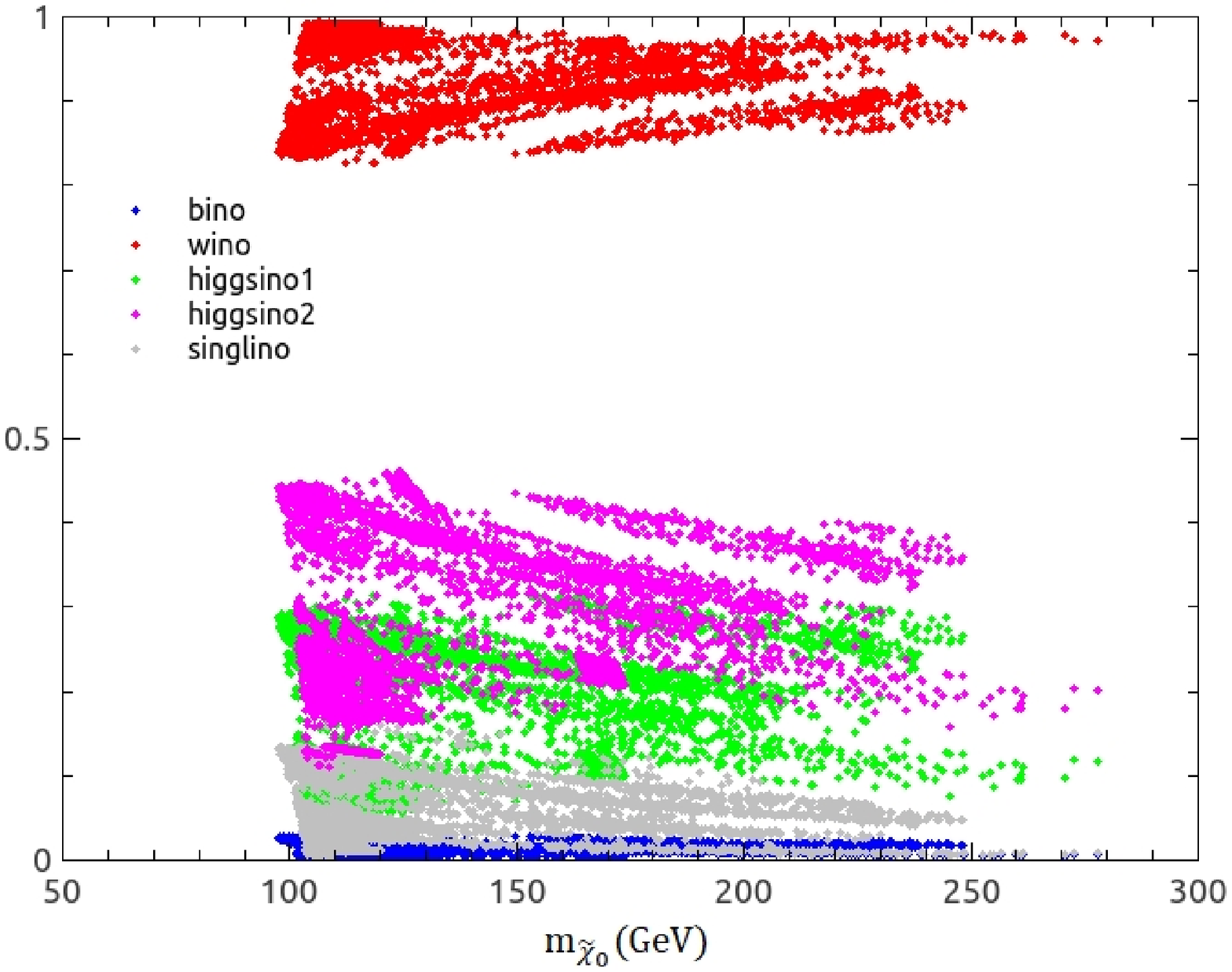}
\includegraphics[width=2.9 in]{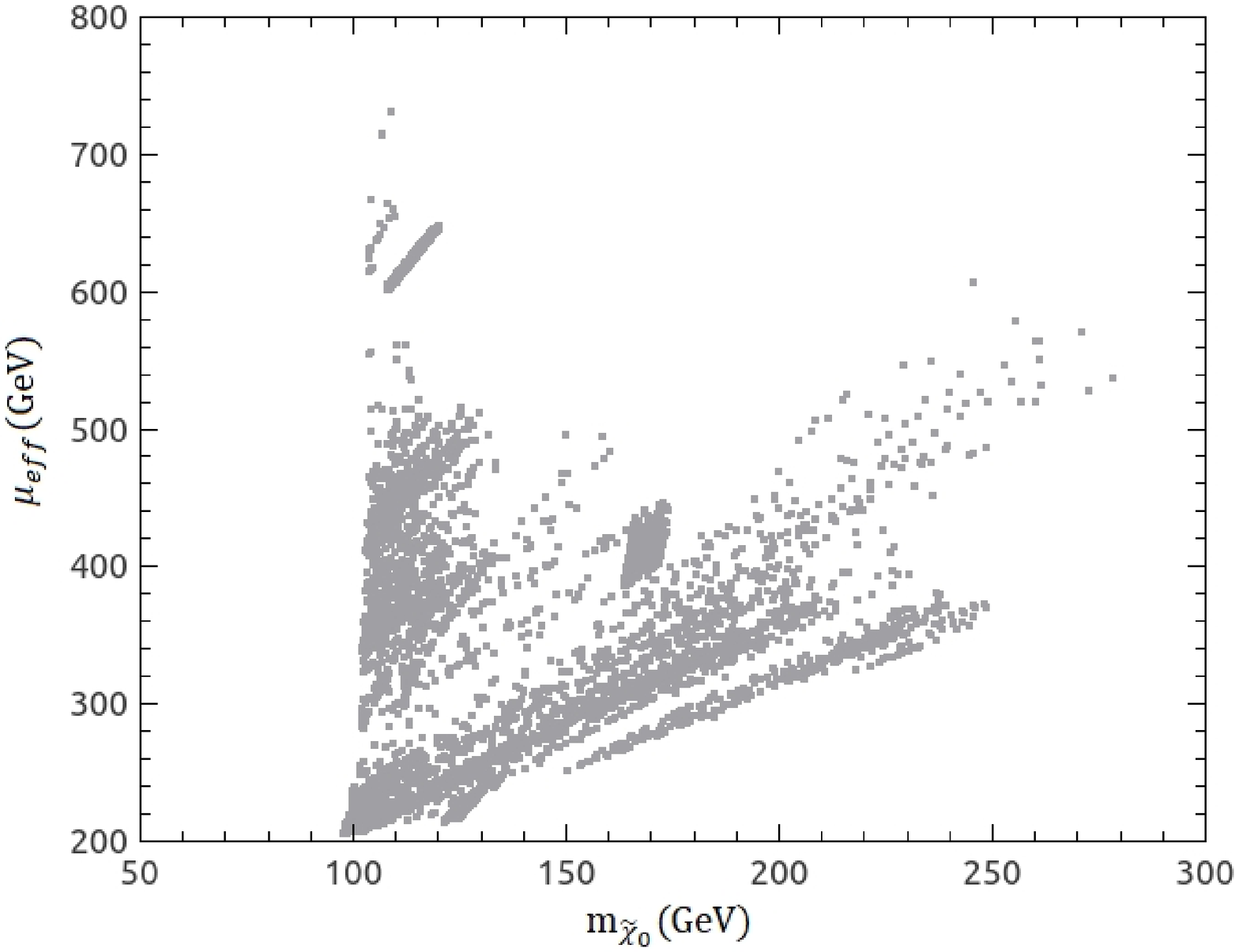}
\end{center}
\caption{The left plot shows the matrix elements to diagonalize the neutralino mass matrix $N_{1i}$ versus the lightest neutralino mass($m_{\tl{\chi}_0^1}$) while the right plot shows the effective $\mu$ parameter for higgsino mass $\mu_{eff}\equiv \la\langle s\rangle$ versus the lightest neutralino mass($m_{\tl{\chi}_0^1}$). All survived points satisfy the LHC and EW precision measurement constraints as well as the required value of $\Delta a_\mu$.}
\label{LSP}
\end{figure}

\begin{figure}[htb]
\begin{center}
\includegraphics[width=4.6 in]{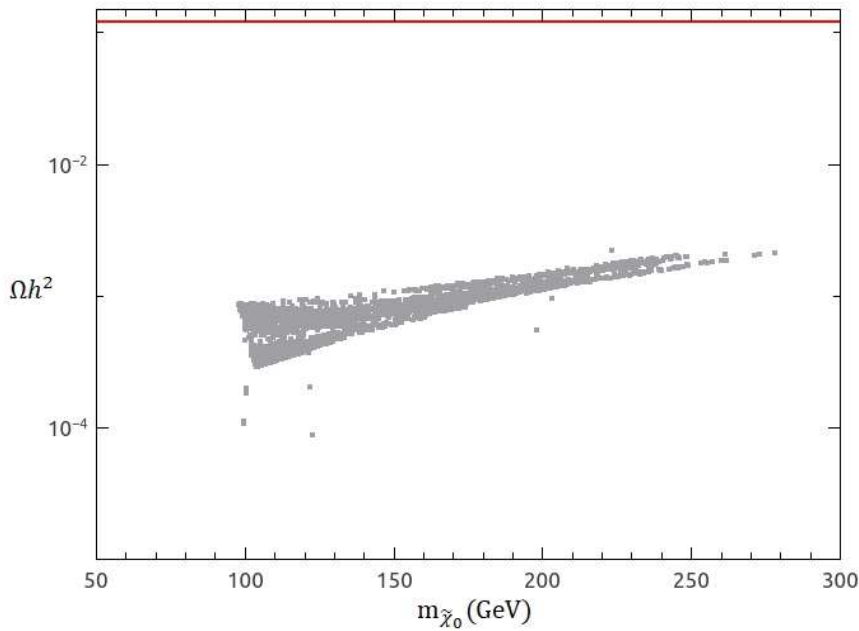}
\end{center}
\caption{The DM relic density versus the DM mass. All points can also satisfy the spin-independent DM direct detection constraints by LUX2016 and PandaX.}
\label{DM}
\end{figure}

\item  The lightest neutralino can act as an excellent DM candidate in the NMSSM. Our numerical scan (see fig.\ref{LSP}) indicates that such neutralino DM mass should lie in the range $100{\rm GeV}<m_{LSP}<300{\rm GeV}$ to account for the $g_\mu-2$ anomaly. The lightest neutralino in the NMSSM should be the mixing of the bino($\tl{B}$), wino($\tl{W}$), higgsino($\tl{H}_{1,2}$) and singlino ($\tl{S}$)
\beqa
\chi_1^0=N_{11}\tl{B}+N_{12}\tl{W}+N_{13}\tl{H}_1+N_{14}\tl{H}_2+N_{15}\tl{S}~,
\eeqa
with $N_{1i}$ the matrix elements to diagonalize the neutralino mass matrix.
  In our scenario, we find that the wino is the dominant component of LSP (see left-plot of fig.\ref{LSP}).
 This result can be partially understood
from the gaugino ratios at the EW scale. The gaugino mass ratios at the weak scale are given approximately by
\beqa
M_1 : M_2 : M_3 \approx [6.6- 3d] : 2 [1-3d] : 6 [-3-3d]~.
\eeqa
 Knowing the range of the deflection parameter $0.45<d<0.85$, the lightest gaugino can be identified from the above formula. The plot for effective $\mu\equiv \la\langle s\rangle$ versus $m_{LSP}$ can be seen in the right plot of fig.\ref{LSP}. We can see that the scale of effective $\mu$ is not too large in compare with $m_{\tl{\chi}_0^1}$. So the higgsino components in neutralino LSP can be sizable. The NMSSM-specific singlino component is negligibly small which will not play an important role in DM annihilation processes. It is well known that wino DM will always lead to under abundance of DM relic abundance unless $m_{\tl{W}}\sim 3 $ TeV. This result holds in our scenario (see fig.\ref{DM}). The wino-like DM can not provide full relic abundance with its mass of order ${\cal O}(100)$ GeV. Other DM components, for example, axion or axino, are necessary to provide enough cosmic DM. We also check that the spin-independent DM direct detection constraints, for example, the LUX2016\cite{LUX2016} and the PandaX\cite{PANDAX}, are satisfied for all survived points. 

\eit
\section{Conclusion}

 We propose to realize the (natural) NMSSM spectrum from deflected AMSB with new messenger-matter interactions. With additional messenger-matter interactions involving ${\bf 10}\oplus{\bf \overline{10}}$ representation messengers, the muon g-2 anomaly can be solved at $2\sigma$ (or $3\sigma$) level with the corresponding gluino mass range $2.8~{\rm TeV}<m_{\tl{g}}<5.4~{\rm TeV}$ (or $2.6~{\rm TeV}<m_{\tl{g}}<7.3~{\rm TeV}$). Besides, our scenario is fairly natural within which the involved fine tuning can be as low as 47. So, in the framework of AMSB-type scenarios, the NMSSM can be advantageous to explain the muon g-2 anomaly in compare with the MSSM.

 \section*{Appendix: Soft SUSY Parameter For the NMSSM}
From the previous analytic expressions, we can obtain the soft SUSY breaking parameters for the NMSSM at the messenger scale.
The gaugino masses are given as
\beqa
M_i=-F_\phi\f{\al_i(\mu)}{4\pi}\(b_i-d\Delta b_i\)~,
\eeqa
with
\beqa
~(b_1~,b_2~,~b_3)&=&(\f{33}{5},~1,-3)~,~~\nn\\
\Delta(b_1~,b_2~,~b_3)&=&(~3,~3,~3).
\eeqa
The trilinear soft terms are given by
\beqa
A_t&=&\f{F_\phi}{16\pi^2}\[\tl{G}_{y_t}-d(\la_{Q,3}^2+\la_{U,3}^2)\]~,\nn\\
A_b&=&\f{F_\phi}{16\pi^2}\[\tl{G}_{y_b}-\la_{Q,3}^2d\]~,\nn\\
A_\tau&=&\f{F_\phi}{16\pi^2}\[\tl{G}_{y_\tau}-\la_{E,3}^2d\]~,\nn\\
A_\la &=&\f{F_\phi}{16\pi^2}\[\tl{G}_{\la}-d\sum\limits_{a=1,2,3}\(6\la_{Q,a}^2+3\la_{U,a}^2+\la_{E,a}^2\)\]~,\nn\\
A_\ka &=&\f{F_\phi}{16\pi^2}\[\tl{G}_{\ka}-3d\sum\limits_{a=1,2,3}\(6\la_{Q,a}^2+3\la_{U,a}^2+\la_{E,a}^2\)\]~,
\eeqa
with
\beqa
\tl{G}_{\la}&=&4\la^2+2\ka^2+3y_t^2+3y_b^2+y_\tau^2-(3g_2^2+\f{3}{5}g_1^2)~,\nn\\
\tl{G}_{\ka}&=&6\la^2+6\ka^2~,\nn\\
\tl{G}_{y_t}&=&\la^2+6y_t^2+y_b^2-(\f{16}{3}g_3^2+3g_2^2+\f{13}{15}g_1^2)~,\nn\\
\tl{G}_{y_b}&=&\la^2+y_t^2+6y_b^2+y_\tau^2-(\f{16}{3}g_3^2+3g_2^2+\f{7}{15}g_1^2)~,\nn\\
\tl{G}_{y_\tau}&=&\la^2+3y_b^2+4y_\tau^2-(3g_2^2+\f{9}{5}g_1^2)~.
\eeqa

Expressions for scalars are rather complicate, they can be parameterized as the sum of each contributions
\beqa
m_{soft}^2=\delta_{d}+\delta_{I}+\delta_G~.
\eeqa
\bit
\item Pure deflected anomaly mediation contribution $\delta_d$ without new yuakwa couplings
\beqa
\delta^d_{{H}_u}~~&=&\f{F_\phi^2}{16\pi^2}\[\f{3}{2}G_2\al^2_2+\f{3}{10}G_1\al^2_1\]
+\f{F_\phi^2}{(16\pi^2)^2}\[\la^2\tl{G}_\la+3y_t^2\tl{G}_{y_t}\]~,~\nn\\
\delta^d_{{H}_d}~~&=&\f{F_\phi^2}{16\pi^2}\[\f{3}{2}G_2\al^2_2+\f{3}{10}G_1\al^2_1\]
+\f{F_\phi^2}{(16\pi^2)^2}\[\la^2\tl{G}_\la+3y_b^2\tl{G}_{y_b}+y_\tau^2\tl{G}_{y_\tau}\]~,~\nn\\
\delta^d_{\tl{Q}_{L;1,2}}&=&\f{F_\phi^2}{16\pi^2}\[\f{8}{3} G_3 \al^2_3+\f{3}{2}G_2\al^2_2+\f{1}{30}G_1\al^2_1\]~,~\nn\\
\delta^d_{\tl{U}^c_{L;1,2}}&=&\f{F_\phi^2}{16\pi^2}\[\f{8}{3} G_3 \al^2_3+\f{8}{15}G_1\al^2_1\]~,~\nn\\
\delta^d_{\tl{D}^c_{L;1,2}}&=&\f{F_\phi^2}{16\pi^2}\[\f{8}{3} G_3 \al^2_3+\f{2}{15}G_1\al^2_1\]~,~\nn\\
\delta^d_{\tl{L}_{L;1,2}}&=&\f{F_\phi^2}{16\pi^2}\[\f{3}{2}G_2\al_2^2+\f{3}{10}G_1\al_1^2\]~,~\nn\\
\delta^d_{\tl{E}_{L;1,2}^c}&=&\f{F_\phi^2}{16\pi^2}\f{6}{5}G_1\al_1^2~,~
\eeqa
with
\beqa
G_i&=&Nd^2+2N d-b_i~,~\nn\\
(b_1,b_2,b_3)&=&(\f{33}{5},1,-3)~.
\eeqa
Here $N=3$ in our scenario.

For the third generation, we need to include the Yukawa contributions
\beqa
\delta^d_{\tl{Q}_{L,3}}&=&\delta^d_{\tl{Q}_{L;1,2}}+F_\phi^2\f{1}{(16\pi^2)^2}\[y_t^2\tl{G}_{y_t}+y_b^2\tl{G}_{y_b}\]~,\nn\\
\delta^d_{\tl{U}^c_{L,3}}&=&\delta^d_{\tl{U}^c_{L;1,2}}+F_\phi^2\f{1}{(16\pi^2)^2}\[2y_t^2\tl{G}_{y_t}\]~,\nn\\
\delta^d_{\tl{D}^c_{L,3}}&=&\delta^d_{\tl{D}^c_{L;1,2}}+F_\phi^2\f{1}{(16\pi^2)^2}\[2y_b^2\tl{G}_{y_b}\]~,~\nn\\
\delta^d_{\tl{L}_{L,3}}&=&\delta^d_{\tl{L}_{L;1,2}}+F_\phi^2\f{1}{(16\pi^2)^2}\[y_\tau^2\tl{G}_{y_\tau}\]~,~\nn\\
\delta^d_{\tl{E}_{L,3}^c}&=&\delta^d_{\tl{E}_{L;1,2}^c}+F_\phi^2\f{1}{(16\pi^2)^2}\[2y_\tau^2\tl{G}_{y_\tau}\]~,~
\eeqa
The pure anomaly contribution to $m_S^2$:
\beqa
\delta^d_{S}&=&\f{F_\phi^2}{(16\pi^2)^2}\[2\la^2\tl{G}_{\la}+2\ka^2\tl{G}_{\ka}\]~.
\eeqa

\item The anomaly-gauge interference mediation part $\delta_I$ from new Yukawa couplings:

The discontinuity of the Yukawa beta functions across the messenger threshold are given as
\beqa
\Delta \tl{G}_{y_t} &=&\la_{Q,3}^2+\la_{U,3}^2~,\nn\\
\Delta \tl{G}_{y_b} &=&\la_{Q,3}^2~,\nn\\
\Delta \tl{G}_{y_\tau} &=&\la_{E,3}^2~,\nn\\
\Delta \tl{G}_{\la} &=&\sum\limits_{a=1,2,3}\[6\la_{Q,a}^2+3\la_{U,a}^2+\la_{E,a}^2\]~~,\nn\\
\Delta \tl{G}_{\ka}&=&3\sum\limits_{a=1,2,3}\[6\la_{Q,a}^2+3\la_{U,a}^2+\la_{E,a}^2\]~~.
\eeqa
 So we have the interference contributions to the soft parameters:
\beqa
2\delta^I_{\tl{Q}_{L,3}}&=&-\f{d F_\phi^2}{(8\pi^2)^2}\[y_t^2\Delta \tl{G}_{y_t}+y_b^2\Delta\tl{G}_{y_b}\]~,\nn\\
~~2\delta^I_{\tl{U}^c_{L,3}}&=&-\f{d F_\phi^2}{(8\pi^2)^2}\[2y_t^2\Delta \tl{G}_{y_t}\]~,\nn\\
~~2\delta^I_{\tl{D}^c_{L,3}}&=&-\f{d F_\phi^2}{(8\pi^2)^2}\[2y_b^2\Delta \tl{G}_{y_b}\]~,\nn\\
~~2\delta^I_{\tl{L}_{L,3}}&=&-\f{d F_\phi^2}{(8\pi^2)^2}\[y_\tau^2\Delta \tl{G}_{y_\tau}\]~,\nn\\
~~2\delta^I_{\tl{E}^c_{L,3}}&=&-\f{d F_\phi^2}{(8\pi^2)^2}\[2y_\tau^2\Delta \tl{G}_{y_\tau}\]~,\nn\\
2\delta^I_{{H}_u}&=&
-\f{d F_\phi^2}{(8\pi^2)^2}\[\la^2\Delta \tl{G}_\la+3y_t^2\Delta \tl{G}_{y_t}\]~,\nn\\
2\delta^I_{{H}_d}&=&-\f{d F_\phi^2}{(8\pi^2)^2}\[\la^2\Delta\tl{G}_\la+3y_b^2\Delta \tl{G}_{y_b}+y_\tau^2\Delta \tl{G}_{y_\tau}\]~,~\nn\\
2\delta^I_{S}&=&-\f{d F_\phi^2}{(8\pi^2)^2}\(2\la^2\Delta\tl{G}_\la+2\ka^2\Delta \tl{G}_{\ka}\)~,~\nn\\
\delta^I_{\tl{Q}_{L;1,2}}&=&0,~~~~~~\delta^I_{\tl{U}^c_{L;1,2}}=0~,~~~~~\delta^I_{\tl{D}^c_{L;1,2}}=0~,~~~\delta^I_{\tl{L}_{L;1,2}}=0~,~
{\delta^I_{\tl{E}_{L;1,2}^c}}=0~.
\eeqa

\item The pure gauge mediation contributions $\delta_G$:
\beqa
\delta_{\tl{Q}_{L,3}}&=&\f{d^2 F_\phi^2}{(16\pi^2)^2}\[\la_{Q,3}^2\(G_{\la_{Q,3}}^+\)\]-\f{d^2 F_\phi^2}{(16\pi^2)^2}\[y_t^2\Delta \tl{G}_{y_t}+y_b^2\Delta\tl{G}_{y_b}\],\nn\\
\delta_{\tl{Q}_{L,2}}&=&\f{d^2 F_\phi^2}{(16\pi^2)^2}\[\la_{Q,a}^2\(G_{\la_{Q,a}}^+\)\]~,~~~~~(a=1,2)~,\nn\\
\delta_{\tl{U}^c_{L,3}}&=&\f{d^2 F_\phi^2}{(16\pi^2)^2}\[\la_{U,3}^2\(G_{\la_{U,3}}^+\)\]-\f{d^2 F_\phi^2}{(16\pi^2)^2}\[2y_t^2\Delta \tl{G}_{y_t}\]~,\nn
\eeqa\beqa
\delta_{\tl{U}^c_{L;a}}&=&\f{d^2 F_\phi^2}{(16\pi^2)^2}\[\la_{U,a}^2\(G_{\la_{U,a}}^+\)\]~,~~~~~(a=1,2)\nn\\
\delta_{\tl{E}_{L,3}^c}&=&\f{d^2 F_\phi^2}{(16\pi^2)^2}\[\la_{E,3}^2\(G_{\la_{E,3}}^+\)\]
-\f{d^2 F_\phi^2}{(16\pi^2)^2}\[2y_\tau^2\Delta \tl{G}_{y_\tau}\],\nn\\
\delta_{\tl{E}_{L,a}^c}&=&\f{d^2 F_\phi^2}{(16\pi^2)^2}\[\la_{E,a}^2\(G_{\la_{E,a}}^+\)\]~,~~~~~(a=1,2)\nn\\
\delta_{\tl{S}}&=&
\f{d^2 F_\phi^2}{(16\pi^2)^2}\sum\limits_{a=1,2,3}\left\{6\[\la_{Q,a}^2\(G_{\la_{Q,a}}^+\)\]
+3\[\la_{U,a}^2\(G_{\la_{U,a}}^+\)\]+\[\la_{E,a}^2\(G_{\la_{E,a}}^+\)\]\right\}~\nn\\
&-&\f{d^2 F_\phi^2}{(16\pi^2)^2}\(2\la^2\Delta\tl{G}_\la+2\ka^2\Delta \tl{G}_{\ka}\)~.\nn\\
\delta_{H_u}~&=&-\f{d^2 F_\phi^2}{(16\pi^2)^2}\[\la^2\Delta \tl{G}_\la+3y_t^2\Delta \tl{G}_{y_t}\]~,\nn\\
\delta_{H_d}~&=&-\f{d^2 F_\phi^2}{(16\pi^2)^2}\[\la^2\Delta\tl{G}_\la+3y_b^2\Delta \tl{G}_{y_b}+y_\tau^2\Delta \tl{G}_{y_\tau}\]~,\nn\\
\delta_{\tl{D}^c_{L,3}}&=&-\f{d^2 F_\phi^2}{(16\pi^2)^2}\[2y_b^2\Delta \tl{G}_{y_b}\]~,\nn\\
\delta_{\tl{L}_{L,3}}&=&-\f{d^2 F_\phi^2}{(16\pi^2)^2}\[y_\tau^2\Delta \tl{G}_{y_\tau}\]~,~\nn\\
\delta_{\tl{D}^c_{L,a}}&=&0~,~~~\delta_{\tl{L}_{L,a}}=0~,~~(a=1,2).
\eeqa
with
\beqa
G_{\la_{Q,3}}^+&=&(\la_X^Q)^2+8\la_{Q,3}^2+7\la_{Q,2}^2+7\la_{Q,1}^2+3(\la_{U,3}^2+\la_{U,2}^2+\la_{U,1}^2)\nn\\
               &+&(\la_{E,3}^2+\la_{E,2}^2+\la_{E,1}^2)+ y_t^2+y_b^2+2\ka^2+2\la^2- \f{16}{3}g_3^2-{3}g_2^2-\f{1}{15}g_1^2~,\nn\\
G_{\la_{Q,2}}^+&=&(\la_X^Q)^2+7\la_{Q,3}^2+8\la_{Q,2}^2+7\la_{Q,1}^2+3(\la_{U,3}^2+\la_{U,2}^2+\la_{U,1}^2)\nn\\
               &+&(\la_{E,3}^2+\la_{E,2}^2+\la_{E,1}^2)+2\ka^2+2\la^2 - \f{16}{3}g_3^2-{3}g_2^2-\f{1}{15}g_1^2~,\nn\\
G_{\la_{Q,1}}^+&=&(\la_X^Q)^2+7\la_{Q,3}^2+7\la_{Q,2}^2+8\la_{Q,1}^2+3(\la_{U,3}^2+\la_{U,2}^2+\la_{U,1}^2)\nn\\
               &+&(\la_{E,3}^2+\la_{E,2}^2+\la_{E,1}^2)+2\ka^2+2\la^2 - \f{16}{3}g_3^2-{3}g_2^2-\f{1}{15}g_1^2~,\nn\\
G_{\la_{U,3}}^+&=&(\la_X^U)^2+6\la_{Q,3}^2+6\la_{Q,2}^2+6\la_{Q,1}^2+5\la_{U,3}^2+4\la_{U,2}^2+4\la_{U,1}^2\nn\\
               &+&(\la_{E,3}^2+\la_{E,2}^2+\la_{E,1}^2)+2\ka^2+2\la^2+2y_t^2 - \f{16}{3}g_3^2-\f{16}{15}g_1^2~,\nn\\
G_{\la_{U,2}}^+&=&(\la_X^U)^2+6\la_{Q,3}^2+6\la_{Q,2}^2+6\la_{Q,1}^2+4\la_{U,3}^2+5\la_{U,2}^2+4\la_{U,1}^2\nn\\
               &+&(\la_{E,3}^2+\la_{E,2}^2+\la_{E,1}^2)+2\ka^2+2\la^2 - \f{16}{3}g_3^2-\f{16}{15}g_1^2~,\nn\\
G_{\la_{U,1}}^+&=&(\la_X^U)^2+6\la_{Q,3}^2+6\la_{Q,2}^2+6\la_{Q,1}^2+4\la_{U,3}^2+4\la_{U,2}^2+5\la_{U,1}^2\nn\\
               &+&(\la_{E,3}^2+\la_{E,2}^2+\la_{E,1}^2)+2\ka^2+2\la^2 - \f{16}{3}g_3^2-\f{16}{15}g_1^2~,\nn\\
\eeqa\beqa
G_{\la_{E,3}}^+&=&(\la_X^E)^2+6\la_{Q,3}^2+6\la_{Q,2}^2+6\la_{Q,1}^2+3(\la_{U,3}^2+\la_{U,2}^2+\la_{U,1}^2)\nn\\
               &+&(3\la_{E,3}^2+2\la_{E,2}^2+2\la_{E,1}^2)+2\ka^2+2\la^2+2y_\tau^2 -\f{12}{5}g_1^2~,\nn\\
G_{\la_{E,2}}^+&=&(\la_X^E)^2+6\la_{Q,3}^2+6\la_{Q,2}^2+6\la_{Q,1}^2+3(\la_{U,3}^2+\la_{U,2}^2+\la_{U,1}^2)\nn\\
               &+&(2\la_{E,3}^2+3\la_{E,2}^2+2\la_{E,1}^2)+2\ka^2+2\la^2 -\f{12}{5}g_1^2~,\nn\\
G_{\la_{E,1}}^+&=&(\la_X^E)^2+6\la_{Q,3}^2+6\la_{Q,2}^2+6\la_{Q,1}^2+3(\la_{U,3}^2+\la_{U,2}^2+\la_{U,1}^2)\nn\\
               &+&(2\la_{E,3}^2+2\la_{E,2}^2+3\la_{E,1}^2)+2\ka^2+2\la^2 -\f{12}{5}g_1^2~,
\eeqa

\eit

\begin{acknowledgments}
We are very grateful to the referee for helpful discussions and efforts to improve our draft. This work was supported by the
Natural Science Foundation of China under grant numbers 11675147,11105124; by the Open Project Program of State Key Laboratory of
Theoretical Physics, Institute of Theoretical Physics, Chinese Academy of Sciences, P.R.
China (No.Y5KF121CJ1); by the Innovation Talent project of Henan Province under grant
number 15HASTIT017.
\end{acknowledgments}


\begin{thebibliography}{99}
\vspace{-1mm}
\bibitem{ATLAS:higgs} G. Aad et al.(ATLAS Collaboration), Phys. Lett. B710, 49 (2012).

\bibitem{CMS:higgs} S. Chatrachyan et al.(CMS Collaboration), Phys. Lett.B710, 26 (2012).

\bibitem{nmssm} U. Ellwanger, C. Hugonie and A. M. Teixeira, Phys. Rept. 496, 1 (2010);\\
                M. Maniatis, Int. J. Mod. Phys. A 25, 3505 (2010).

\bibitem{atlas:gluino}The ATLAS collaboration, ATLAS-CONF-2016-052.

\bibitem{atlas:stop}The ATLAS collaboration, ATLAS-CONF-2016-050.


\bibitem{sm:g-2} K. Hagiwara, R. Liao, A. D. Martin, D. Nomura and T. Teubner, J. Phys. G 38, 085003(2011).

\bibitem{e821} G. W. Bennett et al. [Muon g-2 Collaboration], Phys. Rev. D 73, 072003 (2006).

\bibitem{mssm:g-2} T. Moroi, Phys. Rev. D 53, 6565 (1996).

\bibitem{AMSB}
L.~Randall and R.~Sundrum,
Nucl.\ Phys.\ B {\bf 557}, 79 (1999);
G.~F.~Giudice, M.~A.~Luty, H.~Murayama and R.~Rattazzi,
JHEP {\bf 9812}, 027 (1998).

\bibitem{AMSB:RGE} I. Jack, D.R.T. Jones, Phys.Lett. B465 (1999) 148-154.

\bibitem{deflect} A. Pomarol and R. Rattazzi, JHEP 9905, 013 (1999); \\
    R. Rattazzi, A. Strumia, James D. Wells, Nucl.Phys.B576:3-28(2000).


\bibitem{Fei:1508.01299} Fei Wang, Phys.Lett.B 751(2015)402-407.

\bibitem{NMSSM:GMSB} A. de Gouvea, A. Friedland, H. Murayama, Phys. Rev. D 57 (1998)5676.

\bibitem{NMSSM:AMSB} R. Kitano, G.D. Kribs, H. Murayama, Phys. Rev. D 70 (2004) 035001.

 \bibitem{okada} Nobuchika Okada, Phys.Rev. D65 (2002) 115009;\\
                    Nobuchika Okada, Hieu Minh Tran, Phys.Rev. D87 (2013) 3, 035024.

\bibitem{Fei:1505.02785}Fei Wang, Wenyu Wang, Jin Min Yang, Yang Zhang, JHEP07(2015)138.

\bibitem{wavefunction:hep-ph/9706540} G.F. Giudice, R. Rattazzi, Nucl.Phys.B511:25-44,1998.
\bibitem{chacko} Z. Chacko and E. Ponton, Phys.Rev. D66 (2002) 095004.
\bibitem{Fei:1602.01699} Fei Wang, Jin Min Yang, Yang Zhang, JHEP04(2016)177.
\bibitem{Fei:1703.10894} Fei Wang, Wenyu Wang, Jin Min Yang, arXiv:1703.10894

\bibitem{shih} Jared A. Evans, David Shih, JHEP08(2013)093.




\bibitem{NMSSM:light-stop} U. Ellwanger, JHEP 1203, 044 (2012);\\
                        Z. Kang, J. Li and T. Li, JHEP 1211, 024 (2012);\\
                        S. F. King, M. Muhlleitner and R. Nevzorov, Nucl. Phys. B 860, 207 (2012);\\
                        J. J. Cao, et al., JHEP 1203, 086 (2012), JHEP08(2016)037,JHEP10(2016)136;\\
                        K. Choi, et al., JHEP 1302, 090 (2013);\\
                        S. F. King, et al., Nucl. Phys. B 870, 323 (2013);\\
                        M. Badziak, M. Olechowski and S. Pokorski, JHEP 1306, 043 (2013);\\
                        S. Moretti, S. Munir and P. Poulose, Phys. Rev. D 89, no. 1, 015022 (2014).


\bibitem{ellwanger:0612134} U. Ellwanger and C. Hugonie, Comput. Phys. Commun. 177:399-407, 2007.

\bibitem{NMSSMTOOLS} B. Allanach et al., Comp. Phys. Commun. 180:8-25(2009);\\
                     U. Ellwanger, C.-C. Jean-Louis, A. M. Teixeira, JHEP 0805 (2008) 044.




\bibitem{B-physics} V. Khachatryan et al. [CMS and LHCb Collaborations], Nature 522, 68 (2015);
\bibitem{NMSSM:higgs2loop}M. D. Goodsell, K. Nickel and F. Staub, Phys. Rev. D 91, 035021 (2015).
\bibitem{Planck}  P. A. R. Ade et al. [Planck Collaboration], Astron. Astrophys. 571, A16 (2014).

\bibitem{WMAP}   J. Dunkley et al. [WMAP Collaboration], Astrophys. J. Suppl. 180, 306 (2009).
\bibitem{Mg-2:collaboration} M. Byrne, C. Kolda and J. E. Lennon, Phys. Rev. D 67, 075004 (2003).
\bibitem{muon:g-2} Manfred Lindner, Moritz Platscher, Farinaldo S. Queiroz, arXiv:1610.06587;\\
                   M. Davier, A. Hoecker, B. Malaescu, and Z. Zhang, Eur. Phys. J. C71 (2011)1515;\\
                    A.~Kobakhidze, M.~Talia, L.~Wu, Phys.\ Rev.\ D{\bf 95},055023(2017).
\bibitem{NMSSM:g-2} Florian Domingo, Ulrich Ellwanger,JHEP 0807:079,2008.
\bibitem{ellwanger:1512.04281} Ulrich Ellwanger, Matias Rodriguez-Vazquez, JHEP 1602 (2016)096.
\bibitem{king:1408.1120} S.F. King, M. Muhlleitner, R. Nevzorov, K. Walz, Phys. Rev. D 90, 095014 (2014).
\bibitem{Bomark:2015fga}
  N.~E.~Bomark, S.~Moretti and L.~Roszkowski,
  J.\ Phys.\ G {\bf 43}, no. 10, 105003 (2016).
??
 \bibitem{Djouadi:2008uw}
  A.~Djouadi {\it et al.}, JHEP {\bf 0807}, 002 (2008).

 \bibitem{Cao:2013gba}
  J.~Cao, F.~Ding, C.~Han, J.~M.~Yang and J.~Zhu, JHEP {\bf 1311}, 018 (2013)



\bibitem{moretti:1205.1683}  Mosleh Almarashi, Stefano Moretti, arXiv:1205.1683 [hep-ph].
\bibitem{ellwanger:0503203}  Ulrich Ellwanger, John F. Gunion, Cyril Hugonie, JHEP0507:041,2005.

\bibitem{rnaturalsusy} H. Baer, V. Barger, Peisi Huang, A. Mustafayev, X. Tata, Phys.Rev.Lett. 109 (2012) 161802.

\bibitem{fei-cao}  Junjie Cao, Yangle He, Fei Wang, To appear soon.

\bibitem{LUX2016}  D.~S.~Akerib {\it et al.},
  arXiv:1608.07648 [astro-ph.CO].

\bibitem{PANDAX}  C. Fu et al., Spin-dependent WIMP-nucleon cross section limits from first data of PandaX-II
experiment, Phys. Rev. Lett. 118, 071301 (2017)[arXiv:1611.06553].


\end{thebibliography}
\end{document}